\documentclass[a4paper,11pt]{article}
\pdfoutput=1 

\usepackage{jcappub} 

\usepackage[T1]{fontenc} 
\usepackage[utf8]{inputenc}
\usepackage[normalem]{ulem}
\usepackage{afterpage}
\usepackage{graphicx}
\usepackage{amsmath}
\usepackage{amssymb}
\usepackage{color}

\newcommand{\beq}{\begin{equation}}
\newcommand{\eeq}{\end{equation}}

\newcommand{\pp}{\partial}
\newcommand{\dd}{\mathrm{d}}
\newcommand{\cH}{\mathcal{H}}

\newcommand{\lf}{{\sc lat}field{\sc 2}}

\title{\boldmath Relativistic N-body simulations with massive neutrinos}

\author[a]{Julian~Adamek,}
\author[b]{Ruth~Durrer}
\author[b]{and Martin~Kunz}

\affiliation[a]{Laboratoire Univers et Th\'eories, Observatoire de Paris -- PSL Research University -- CNRS -- Universit\'e Paris Diderot -- Sorbonne Paris Cit\'e, 5 Place Jules Janssen, 92195 Meudon CEDEX, France}
\affiliation[b]{D\'epartement de Physique Th\'eorique \& Center for Astroparticle Physics, Universit\'e de Gen\`eve, 24 Quai E.\ Ansermet, 1211 Gen\`eve 4, Switzerland}

\emailAdd{julian.adamek@obspm.fr}
\emailAdd{ruth.durrer@unige.ch}
\emailAdd{martin.kunz@unige.ch}

\abstract{Some of the dark matter in the Universe is made up of massive neutrinos. Their impact on the formation
of large scale structure can be used to determine their absolute mass scale from cosmology, but to this end accurate numerical
simulations have to be developed. Due to their relativistic nature, neutrinos pose additional challenges when one tries to include them
in N-body simulations that are traditionally based on Newtonian physics. Here we present the
first numerical study of massive neutrinos that uses a fully relativistic approach. Our N-body code, \textit{gevolution}, is based on a
weak-field formulation of general relativity that naturally provides a self-consistent framework for relativistic particle species. This
allows us to model neutrinos from first principles, without invoking any ad-hoc recipes. Our simulation suite comprises some of the
largest neutrino simulations performed to date. We study the effect of massive neutrinos on the nonlinear power spectra and the halo mass
function, focusing on the interesting mass range between 0.06~eV and 0.3~eV and including a case for an inverted mass hierarchy.}

\begin{document}
\maketitle
\flushbottom

\section{Introduction}
\label{sec:intro}

Since the discovery of the Higgs particle at the LHC \cite{Chatrchyan:2012xdj,Aad:2012tfa} the standard model of particle
physics is formally complete, except for the still unknown physical origin of the neutrino masses. Neutrino mass differences have
been measured through neutrino oscillation experiments (e.g.\ \cite{Kajita:2016cak,McDonald:2016ixn,Abe:2011fz,Aguilar:2001ty,Mention:2011rk,Aguilar-Arevalo:2013pmq,Feldman:2013vca}), which
indicate squared mass differences of $\Delta m_{21}^2 =$ (7.37$\pm$0.18) $\times$ 10$^{-5}\, \mathrm{eV}^2$ and $|\Delta m^2| =$ (2.50 $\pm$ 0.05) $\times$ 10$^{-3}\, \mathrm{eV}^2$ (for details and the definition of $\Delta m^2$ see section 14 of \cite{Olive:2016xmw}).
But the absolute neutrino mass scale is still unknown, and its determination may provide important clues to the last
remaining gap in the standard model of particle physics.

In addition, massive neutrinos impact structure formation in the Universe, e.g.\ \cite{Lesgourgues:2006nd}. At early times
neutrinos are relativistic and contribute to the radiation density. As the Universe expands, massive neutrinos become
non-relativistic and contribute to the total dark matter. But due to their initially high velocities they  free-stream
out of over-densities and thus reduce the clustering of the dark matter, relative to the standard $\Lambda$-cold-dark-matter
($\Lambda$CDM) model with massless neutrinos. This impact of neutrino masses on the galaxy clustering in the Universe offers
upcoming large galaxy and weak lensing surveys like DESI \cite{Levi:2013gra}, Euclid \cite{Laureijs:2011gra,Amendola:2012ys},
LSST \cite{Abell:2009aa,Abate:2012za} or SKA \cite{Carilli:2004nx} the opportunity to determine the absolute neutrino mass 
scale with cosmological observations. The \textit{Planck} measurement of the cosmic
microwave background alone already limits the sum of neutrino masses to $\sum m_\nu <$ 0.49~eV at 95\% confidence level \cite{Ade:2015xua}.
Adding external data, mainly baryon acoustic oscillations, results in $\sum m_\nu <$ 0.23~eV, and the most stringent cosmological constraint, using the Ly-$\alpha$ forest, is $\sum m_\nu <$ 0.14~eV at 95\% \cite{Palanque-Delabrouille:2014jca}. Current laboratory limits from measurements
of the $\beta$-decay kinematics are about an order of magnitude higher, see~\cite{Olive:2016xmw}, and will reach a sensitivity
comparable to the \textit{Planck} constraints only with the KATRIN experiment (e.g.\ \cite{Drexlin:2013lha}).

However, in order to use cosmology to measure the neutrino masses we have to be able to predict their impact on clustering
to high accuracy on a wide range of scales, including those where perturbation theory no longer applies. Mistakes in the
predictions will not only bias a measurement of the neutrino masses but generally induce a large systematic error in the
results of those surveys, which would compromise the goals of precision cosmology. For this reason we have to rely at least
partially on numerical simulations of galaxy clustering including massive neutrinos, as there is no other way to study the
evolution of perturbations on strongly non-linear scales. But while cold dark matter (CDM) lives on an effectively three-dimensional
sheet in the six-dimensional phase space due to its small velocity dispersion, the relativistic neutrinos fill the full
six-dimensional volume, hugely reducing the particle density in phase space and thus increasing the shot noise in any quantity
derived from the simulations. An additional problem, particularly for Newtonian simulations, stems from the high initial
momenta that require either a relativistic treatment or else lead to superluminal propagation speeds for the neutrinos.

In this paper we discuss simulations of cosmological structure formation including massive neutrinos with the relativistic
N-body code \textit{gevolution} \cite{Adamek:2015eda}. We start with a description of the numerical implementation and the
differences to Newtonian simulations of structure formation with neutrinos in Sections~\ref{sec:numerics} and \ref{sec:newtonian}.
We present our simulation suite in Section~\ref{sec:sims} and discuss
results on the matter power spectrum in Section~\ref{sec:matterspec}. We study relativistic spectra that are not
available in Newtonian simulations in Section~\ref{sec:relspec} and  the halo mass function in Section~\ref{sec:hmf}.
Section~\ref{sec:conclusions} summarizes our conclusions. In Appendix~\ref{app:integrator} we briefly discuss how to
implement a symplectic leapfrog integrator for relativistic particles. Some convolution integrals for second-order source terms of the metric perturbations are collected in Appendix~\ref{app:2order}.

\section{Numerical approach}
\label{sec:numerics}

In this section we discuss the key elements of our relativistic N-body code \textit{gevolution}, with particular
emphasis on the features relevant for simulating relativistic species. These features are available with the latest public
release of the code, which can be obtained at \url{https://github.com/gevolution-code/gevolution-1.1.git}. While aiming
for a self-contained presentation, we refer the reader to \cite{Adamek:2014xba,Adamek:2016zes,Adamek:2017grt} for additional details.

\subsection{Weak-field expansion of the metric}

The relativistic framework employed in \textit{gevolution} is based on a geometric interpretation of gravity. The gravitational
fields, sourced by the stress-energy tensor, act through a deformation of spacetime that modifies geodesic motion. This has to be
contrasted to the Newtonian picture of a two-body force. On cosmological scales the gravitational fields are extremely weak, as
can be seen easily by considering e.g.\ the compactness parameter of the structure at those scales:\ galaxies and clusters
have a typical size many orders of magnitude larger than their Schwarzschild radius. We can therefore choose an appropriate ansatz
for the perturbed metric and perform a systematic expansion of Einstein's field equations in terms of the weak gravitational fields. However, it
is important to pick a coordinate system in which the geometry indeed remains weakly perturbed, and some of the simple coordinate choices
can fail miserably. We employ the coordinate system of the Poisson gauge that has proven to be very well adapted to this requirement
(see also \cite{Brown:2009tg}). The line element reads
\begin{equation}
 ds^2 = g_{\mu\nu} dx^\mu dx^\nu = a^2(\tau) \left[-e^{2\psi} d\tau^2 - 2 B_i dx^i d\tau + \left(e^{-2\phi} \delta_{ij} + h_{ij}\right) dx^i dx^j\right] \, ,
\end{equation}
where $\tau$ is conformal time, $x^i$ are coordinates on the spacelike hypersurface, and $\psi$, $\phi$, $B_i$, $h_{ij}$
are the gravitational fields that characterize the deformation of the geometry away from a Friedmann-Lema\^itre model.
The coordinate system is fixed by the gauge conditions
\begin{equation}
 \delta^{ij} B_{i,j} = \delta^{ij} h_{ij} = \delta^{jk} h_{ij,k} = 0 \, ,
\end{equation}
where we introduced the shorthand $f_{,i} \doteq \pp f / \pp x^i$. We will also use $f' \doteq \pp f / \pp \tau$ and
$\Delta f \doteq \delta^{ij} f_{,ij}$, as well as $\cH \doteq a'/a$.

Note that the two scalar potentials $\phi$ and $\psi$ are introduced in a slightly different way than the corresponding potentials
$\Phi$ and $\Psi$ used in the original version of \textit{gevolution} \cite{Adamek:2016zes}. This change of convention pays tribute to
the fact that $(\phi - \psi)$ has an improved behaviour at second order compared to $(\Phi - \Psi)$ \cite{Adamek:2017grt}. At first
order the two definitions are equivalent, and we will continue using the symbol $\chi \doteq \phi - \psi$ which is now defined
with respect to the new convention.

In the linear evolution of standard cosmology the vector mode $B_i$ has no source, it decays and is usually discarded.
The spin-2 field $h_{ij}$ and the gravitational slip $\chi$ can only be sourced by relativistic species at first order, and
hence they also decay inside the horizon once radiation domination comes to an end some 100\,000 years after the Big Bang.
Therefore, except for the case of a large contribution of primordial gravitational waves, the second-order contributions
to $h_{ij}$ and $\chi$ eventually dominate at (sufficiently large) sub-horizon scales.  A consistent computation of these contributions
includes some quadratic terms constructed from the first-order perturbations $\psi$ and $\phi$. Apart from these terms,
which are only relevant in the regime where stress-energy perturbations are small (otherwise these quadratic weak-field terms
cannot compete with the stress-energy and are subdominant), it is sufficient to truncate the weak-field expansion at leading order.
Einstein's field equations can then be written as
\begin{equation}
 \label{eq:00}
 \left(1 + 2 \phi\right) \Delta \phi - 3 \cH \phi' - 3 \cH^2 \left(\phi - \chi\right) - \frac{1}{2} \delta^{ij} \phi_{,i} \phi_{,j} = -4 \pi G a^2 \delta T^0_0 ,
\end{equation}
\begin{equation}
 \label{eq:B}
 \Delta^2 B_i = 16 \pi G a^2 P_i^j T_j^0\, ,
\end{equation}
\begin{equation}
 \label{eq:chi}
 \Delta^2 \chi - \left(3 \delta^{ik} \delta^{jl} \frac{\pp^2}{\pp x^k \pp x^l} - \delta^{ij} \Delta\right) \phi_{,i} \phi_{,j} = 
  4 \pi G a^2 \left(3 \delta^{ik} \frac{\pp^2}{\pp x^j \pp x^k} - \delta^{i}_j \Delta\right) T_i^j\, ,
\end{equation}
\begin{equation}
 \label{eq:h}
 \Delta^2 \left(h_{ij}'' + 2 \cH h_{ij}' - \Delta h_{ij}\right) - 4 \left(P_i^k P_j^l -\frac{1}{2} P_{ij} P^{kl}\right) \phi_{,k} \phi_{,l} = 16 \pi G a^2 \left(P_{ik} P_j^l -\frac{1}{2} P_{ij} P_k^l\right) T^k_l \, ,
\end{equation}
where we define the transverse projection operator as
\begin{equation}
 P_{ij} \doteq \frac{\pp^2}{\pp x^i \pp x^j} - \delta_{ij} \Delta \, .
\end{equation}
Three additional equations are not written here: a scalar equation obtained from the spatial trace, the longitudinal part of the
momentum constraint, and the spin-1 projection on the spatial hypersurface. These are degenerate with the covariant conservation of
stress-energy which also amounts to two scalar and one spin-1 equations, but one can use them to check this conservation law
within the numerical scheme. Furthermore, we have subtracted a background contribution from the Hamiltonian constraint, given by
the Friedmann equation
\begin{equation}
 \frac{3}{2} \cH^2 = -4 \pi G a^2 \bar{T}^0_0 \, .
\end{equation}
It is often assumed that this subtraction ensures that the remaining perturbations have zero average, but it turns out that
this is a nontrivial requirement that is not automatically fulfilled beyond leading order. Instead of imposing this condition,
it is however sufficient to subtract an approximate background model, i.e.\ a contribution that allows for a finite residual that is
however perturbatively small. In practice we use a reference Friedmann model for the given cosmology and subtract exactly the same
quantity on either side of the equation, hence defining $\delta T^0_0 \doteq T^0_0 - \bar{T}^0_0$ without assuming an exactly
vanishing average for $\delta T^0_0$. Consistency of the equations is dynamically maintained by allowing for a homogeneous mode in $\phi$ and $\psi$.
The scheme remains valid as long as this homogeneous mode remains sufficiently small.

The stress-energy for the particle ensemble is computed non-perturbatively, however, on the weakly perturbed geometry. This
means that the quantities obtained from the particle-mesh projection are dressed by linear geometric corrections that describe,
for instance, the perturbation of the volume element.

\subsection{Canonical momentum and geodesic equation}

Adhering to its general relativistic approach \textit{gevolution} uses a canonical momentum $q_i$, rather than a peculiar velocity,
in order to describe phase space. The advantage of this description is that it remains valid for arbitrarily large momenta,
including in particular the ultra-relativistic limit $q^2 \gg m^2 a^2$, where $m$ is the proper rest-mass. The geodesic equation reads
\begin{equation}
\label{eq:geodesic}
 q_i' = -\frac{2 q^2 + m^2 a^2}{\sqrt{q^2 + m^2 a^2}} \phi_{,i} + \sqrt{q^2 + m^2 a^2} \chi_{,i} - q^j B_{j,i} + \frac{1}{2} \frac{q^j q^k h_{jk,i}}{\sqrt{q^2 + m^2 a^2}} \, .
\end{equation}
The first two terms are the usual gravitational deflection, featuring the correct ultra-relativistic limit relevant for weak lensing.
The third term incorporates frame-dragging which is already a very small effect under most realistic circumstances. The last term
describes the scattering of particles on gravitational waves. This interaction is so weak that it certainly cannot affect structure formation
unless one constructs very exotic scenarios. Therefore, to reduce computational cost, this term is currently neglected in \textit{gevolution}
even though it would be straightforward to include it. Frame-dragging, however, is taken into account.

The coordinate three-velocity is related to $q_i$ as
\begin{equation}
\label{eq:velocity}
 \delta_{ij} \frac{\pp x^j}{\pp \tau} = \frac{q_i}{\sqrt{q^2 + m^2 a^2}} \left[1 + \left(3 - \frac{q^2}{q^2 + m^2 a^2}\right) \phi - \chi + \frac{1}{2} \frac{q^j q^k h_{jk}}{q^2 + m^2 a^2}\right] + B_i - \frac{h_{ij} q^j}{\sqrt{q^2 + m^2 a^2}}\, ,
\end{equation}
where, as above, we may choose to neglect the terms involving $h_{ij}$. The last two equations are directly implemented in the staggered
leapfrog integrator for the particle evolution. Some theoretical aspects of this integrator are discussed in Appendix \ref{app:integrator}.

\subsection{Non-cold dark matter -- N-body versus Boltzmann hierarchy}

Following the terminology used in \textit{CLASS} \cite{Blas:2011rf} we will call
any relativistic species that can be modelled as collisionless massive particles a non-cold dark matter (NCDM) component,
of which massive neutrinos in the late Universe are just one particular realization. The lack of a scattering process
means that such a NCDM component is not driven towards an equilibrium and therefore develops a complex phase-space structure
that cannot be mapped to a fluid description even at first order. In the linear Boltzmann approach such a component is
modelled by taking into account a large number of multipoles in the distribution function and solving their coupled evolution
\cite{Lesgourgues:2011rh}. If high accuracy is required this can easily involve hundreds of coupled equations. The fundamental
limitation of this approach is the assumed linearity: non-linear clustering of NCDM cannot be taken into account easily.

The alternative is to directly use the N-body scheme as a Monte Carlo method to sample the phase space of NCDM. While conceptually
straightforward there are practical limitations on the number of samples one can draw in a simulation, typically leading to very noisy
representations. Relativistic species have a very broad momentum distribution -- this is exactly what distinguishes NCDM from CDM.
One then has to be very careful that the shot noise from sampling does not produce spurious clustering in the simulation. This
issue is greatly alleviated for simulations of massive neutrinos in the context of $\Lambda$CDM since in this case the NCDM makes up
only a small fraction of the dark matter. It is therefore possible to reduce the NCDM shot noise well below the perturbation amplitude
of CDM, at which point it causes no harm anymore. Note that this would not apply in situations where none of the dark matter is cold,
like e.g.\ in warm dark matter models.

We thus have two entirely different approaches at our disposal (see Figure~\ref{f:snapshot} for an illustration) -- the Boltzmann
approach that works in Fourier space and gives very accurate results that are however confined to the linear regime, and the N-body approach
that can follow the evolution in configuration space nonperturbatively but has to deal with shot noise. Several attempts have been made to
combine or extend these two approaches \cite{Brandbyge:2008js,Brandbyge:2009ce,AliHaimoud:2012vj,Banerjee:2016zaa}. In this work we employ
the N-body approach, i.e.\ a separate N-body ensemble is simulated for each NCDM species. However, it is also possible to link \textit{CLASS}
to \textit{gevolution}, thus allowing the code to compute any linear transfer function by calling \textit{CLASS} at runtime. One can then choose
whether one wants to use the linear realization of the perturbations as given by the Boltzmann solution (as was advocated in
\cite{Brandbyge:2008js}), or the noisy but nonperturbative realization of the perturbations as given by the N-body ensemble, as input for
the stress-energy tensor which is used for solving the metric. The choice can be made separately for each species, and in
\textit{gevolution} the user can specify redshift values at which the code switches between the two methods. This can be useful because
the noise is of increasing concern at higher redshift when the CDM perturbations are still rather small and the NCDM particles have large
velocities. At the same time the perturbations are still linear to good approximation and the Boltzmann approach is well justified.

\afterpage{
\begin{figure}[p!]
\begin{center}
\includegraphics[width=\textwidth]{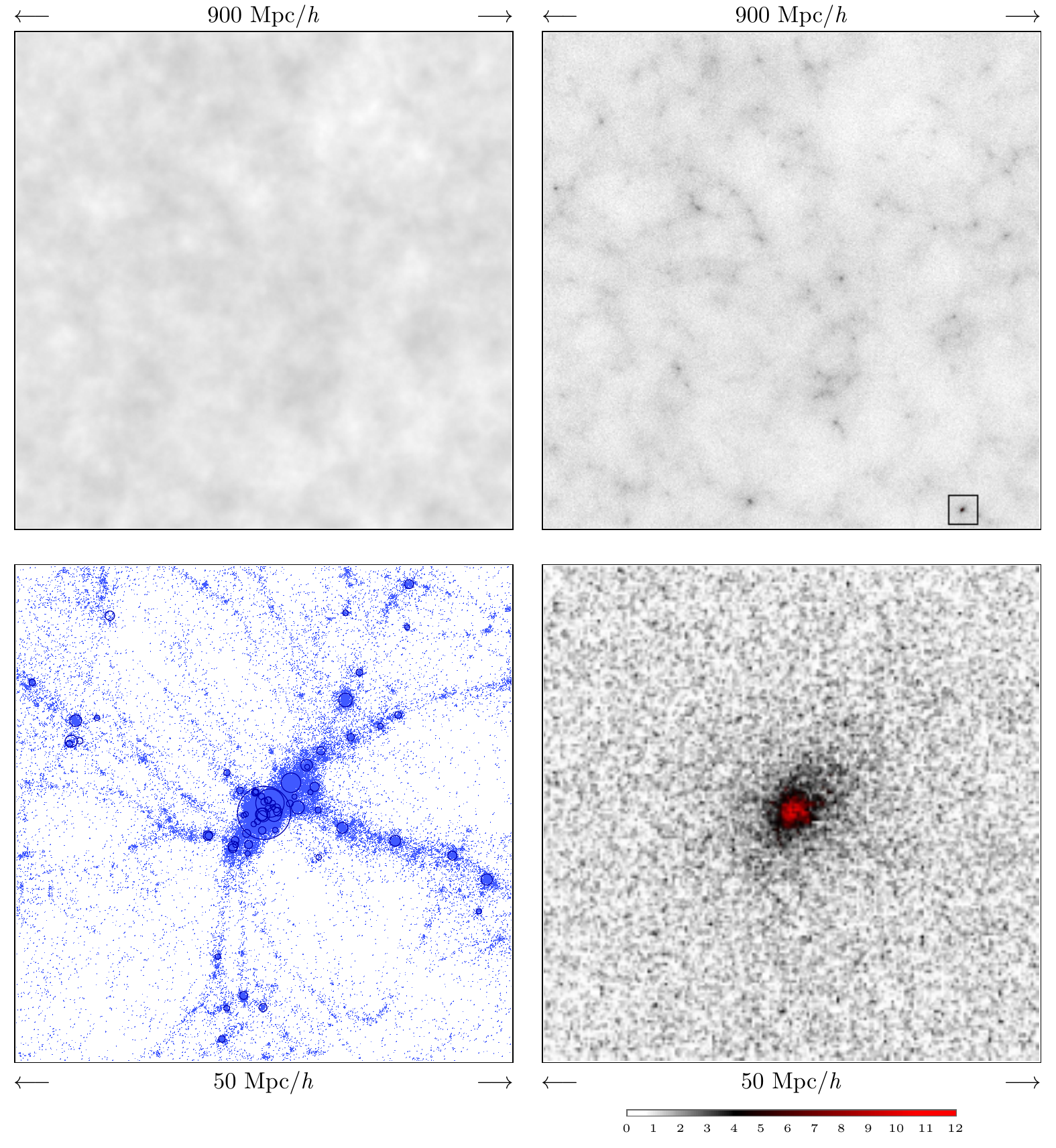}
\end{center}
\caption{\label{f:snapshot}Slice through a simulation with massive neutrinos ($\sum m_\nu = 0.2$ eV) at redshift $z = 0$. The upper
panels show the neutrino density field across the entire simulation box, projected along the $5$~Mpc/$h$ thickness of the slice. The upper
left shows the result obtained using the linear realization of the density field constructed from the transfer functions that were computed
by \textit{CLASS}. The upper right shows the same realization using the N-body method. While being somewhat noisy, the N-body method is able
to follow the nonlinear evolution of the neutrino distribution. The lower right panel is a zoom into a region containing a massive
structure, showing that the density contrast $\delta_\nu = \delta \rho_\nu / \bar{\rho}_\nu$ of the neutrino component can be larger than
unity (scale bar indicates $\delta_\nu + 1$). To provide some context, the CDM distribution for the same region is shown in the lower
left panel, with circles indicating the locations and virial radii of friends-of-friends halos detected with the \textit{ROCKSTAR} halo finder
\cite{Behroozi:2011ju}. Note that the particle snapshots used for this figure were downgraded by a factor of $32$ to make the files more
easy to handle; the actual simulation therefore had a much lower noise level.}
\end{figure}
\clearpage
}

\subsection{Initial conditions}
\label{subsec:IC}

Setting initial conditions for N-body simulations with neutrinos has been considered somewhat of an art. This has a lot to do
with the fact that Newtonian codes have to be repurposed for a task to which they are not very well adapted \cite{Zennaro:2016nqo}.
With \textit{gevolution} the story is quite different: its underlying relativistic framework allows an implementation that is
conceptually clean and straightforward, based on a procedure outlined in \cite{Ma:1993xs}.

We initialize the neutrino N-body ensembles at a redshift where the particle horizon is of the order of the spatial resolution
of the simulation, such that all modes are initially outside the horizon. For our values of the resolution this typically
requires a redshift well above $10^5$. At this time the perturbations are still adiabatic and
Gaussian. Furthermore, the phase-space distribution function is still close to the initial Fermi-Dirac distribution produced by
thermal freeze-out. The initial neutrino momenta are drawn from this initial distribution using a fast rejection-sampling method,
including the temperature monopole and dipole perturbations caused by density and velocity perturbations of the primordial plasma.
The higher moments of the distribution function are negligible far outside the horizon and only build up over time.

\begin{figure}[t]
\begin{center}
\includegraphics[width=\textwidth]{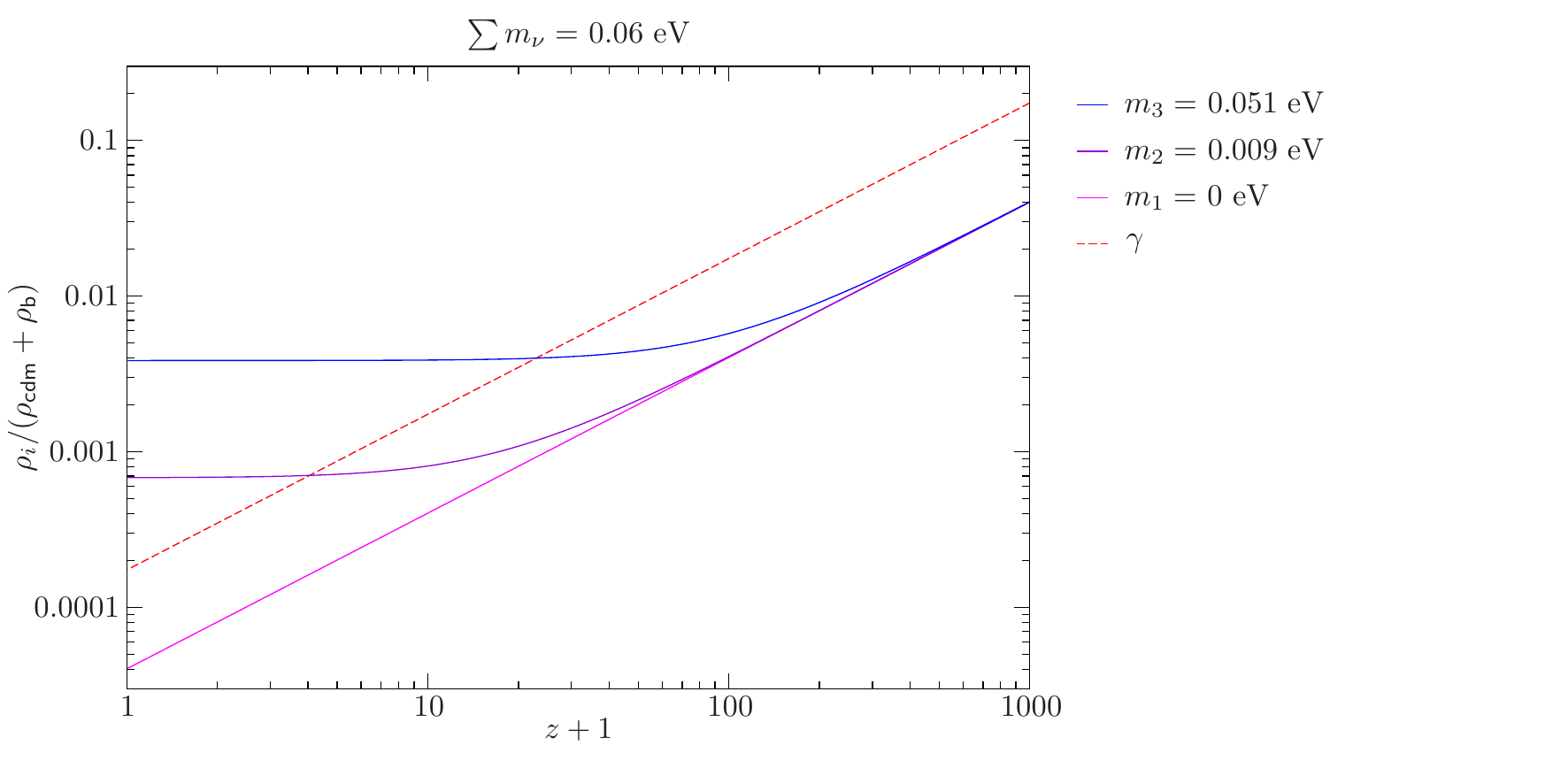}
\end{center}
\caption{\label{f:rhonu}Evolution of the energy density of photons (dashed line) and the three neutrino species in a minimal
mass scenario, relative to the energy density of the remaining matter (CDM and baryons).}
\end{figure}

The free-streaming neutrino particles are then evolved down to a much lower redshift where baryons and CDM will be added.
In all our simulations this happens at $z_\mathrm{ini} \simeq 127$. Following \cite{Ma:1993xs} we
evolve the neutrinos in the linearly perturbed metric, where we obtain the relevant transfer functions of $\phi$ and $\psi$ by
calling \textit{CLASS} at runtime. This procedure is in some sense equivalent to solving the full Boltzmann hierarchy for the neutrino phase space.
At $z_\mathrm{ini}$ we set up the N-body ensemble of CDM and baryons, represented as a single species with perturbations given as
weighted averages of the linear CDM and baryon density and velocity transfer functions. All the baryonic physics happening before photon
decoupling is hence taken into account in these initial conditions. As shown in Figure~\ref{f:rhonu} the relativistic energy density of
each neutrino species has also dropped to $\lesssim 1\%$ of the CDM density at that redshift, allowing the effect of shot noise to be
kept under control.

From this point onward the metric is solved using the N-body ensembles and the simulation eventually proceeds into the
nonperturbative regime. The perturbations of photons, which account for a waning percentage of the total energy density
of the Universe, are neglected and we only keep their contribution to the background. The same is done for any massless neutrino
states. Inside the horizon the perturbations of these radiation components decay and can no longer compete with the growing matter
perturbations. Close to the horizon, on the other hand, their perturbations can still have a subtle effect reaching deep into the
matter dominated era \cite{Brandbyge:2016raj,Adamek:2017grt}. This effect can be taken into account in \textit{gevolution}, again
by constructing the linear perturbations according to transfer functions obtained through \textit{CLASS}. Details on this procedure are
given in \cite{Adamek:2017grt}, but after running some tests we convinced ourselves that we do not need to include it in our production
runs with a box size of 2 Gpc$/h$ and less.

\section{Relation to Newtonian simulations}
\label{sec:newtonian}

N-body simulations with massive neutrinos have so far mostly been carried out using Newtonian codes. We therefore want to
discuss how results should be compared between the different approaches. One challenging aspect is the fact that space and time
are absolute concepts in Newton's theory. This poses the problem of identifying the foliation of spacetime that best corresponds
to the Newtonian picture. The Newtonian gauge, which is just another name for the scalar sector of the Poisson gauge, seems to be
a good choice on scales much smaller than the horizon. In this limit eq.~(\ref{eq:00}) becomes
\begin{equation}
\label{eq:N00}
 \Delta \phi = 4 \pi G a^2 \delta\rho\, ,
\end{equation}
keeping the leading weak-field order only since the next order is already post-Newtonian. The Newtonian approximation is only
valid if all gravitating matter moves at small velocities such that kinetic energy can always be neglected against rest mass energy.
Massive neutrinos fulfill this criterion only to a limited extent, and for the comparison discussed below we will choose neutrino
masses such that kinetic energy is subdominant within the redshift interval covered by the N-body simulation. In the nonrelativistic
limit $q^2 \ll m^2 a^2$ the geodesic equation (\ref{eq:geodesic}) reduces to
\begin{equation}
\label{eq:Ngeodesic}
 q_i' = - m a \phi_{,i} \,,
\end{equation}
where we used $\chi = 0$ as a result of the Newtonian approximation. With eq.~(\ref{eq:velocity}) the peculiar velocity is simply
\begin{equation}
\label{eq:Nvelocity}
 \delta_{ij} \frac{\pp x^j}{\pp \tau} = \frac{q_i}{m a}
\end{equation}
at leading order. Eqs.~(\ref{eq:N00})--(\ref{eq:Nvelocity}) of course are exactly the equations of Newtonian gravity with
$\phi$ playing the role of the Newtonian potential.

The direct link between Newtonian gravity and the Newtonian gauge does unfortunately not persist on very large scales. As one
approaches the cosmological horizon, terms like $\mathcal{H}^2 \phi$ become important in the Hamiltonian constraint (\ref{eq:00})
such that eq.~(\ref{eq:N00}) is no longer a useful truncation. This issue led to an interesting debate on how one should interpret
Newtonian simulations at those scales \cite{Chisari:2011iq,Green:2011wc,Flender:2012nq,Rigopoulos:2013nda,Milillo:2015cva}. For linear
scales the problem has finally been solved in \cite{Flender:2012nq,Fidler:2015npa,Fidler:2016tir} by realizing that there exists a class of
gauges that is compatible with Newtonian dynamics by construction. While this allows one to obtain a relativistic interpretation of
the two-point function of Newtonian simulations at very large separations, many aspects still remain to be worked out. For instance,
it is not obvious how the analysis of weak gravitational lensing would have to be adjusted.

For the purpose of our comparison we will make use of the so-called N-body gauge, introduced in \cite{Fidler:2015npa}. This gauge,
defined in the linear regime, has the property that the equations reduce to the Newtonian ones on arbitrary scales if all sources
of perturbations are non-relativistic. This approximation is excellent for CDM and baryons and works still fairly well for massive
neutrinos as soon as they are sufficiently cold.

In order to allow for a comparison that is minimally affected by aspects of implementation we employ the Newtonian mode of
\textit{gevolution} to run Newtonian simulations. In this case the code first prepares the initial conditions in exactly the same way as
it is done for relativistic runs, explained in Section \ref{subsec:IC}, and then applies an additional step:
from the linear transfer functions provided by \textit{CLASS} the code computes the gauge transformation between Poisson and N-body gauge, i.e.\
two scalar potentials whose gradients are the displacement and velocity correction describing the active coordinate transformation
that has to be applied to the N-body ensemble. After the initial data has thereby been put into N-body gauge the simulation is performed
using the Newtonian equations (\ref{eq:N00})--(\ref{eq:Nvelocity}).

It is worth noting that the neutrino particles in the tail of the distribution move at superluminal velocity under this prescription, as the thermal velocity is assumed to redshift as $1/a$ at all times without accounting for the Lorentz factor.
This issue, mainly relevant at high redshift, has so far been treated as an acceptable limitation of Newtonian simulations. Concerned
by the apparent loss of causality, in \cite{Adamek:2016zes} we advocated to restore Lorentz invariance by replacing
equations (\ref{eq:Ngeodesic}) and (\ref{eq:Nvelocity}) by
\begin{equation}
\label{eq:SRgeodesic}
 q_i' = -\frac{2 q^2 + m^2 a^2}{\sqrt{q^2 + m^2 a^2}} \phi_{,i} \,,
\end{equation}
and
\begin{equation}
\label{eq:SRvelocity}
 \delta_{ij} \frac{\pp x^j}{\pp \tau} = \frac{q_i}{\sqrt{q^2 + m^2 a^2}} \,,
\end{equation}
respectively, without modifying the computation of $\phi$. The latter is probably justified because the relativistic tail of the
neutrino distribution only gives a tiny contribution to the total gravitational field which is dominated by CDM and baryons. It
should be straightforward to implement the above equations of motion in any Newtonian code, and we will call the resulting scheme
``Newtonian gravity with special relativity''.

We therefore have three different simulation types to compare, corresponding to three different levels of approximation. The least accurate
one, which at the same time is the one widely used in the literature, is Newtonian gravity with Newtonian mechanics. The second one, aiming at a
rectification of the evolution of relativistic test particles, is Newtonian gravity with special relativity. The third one, finally, is
our general relativistic weak-field framework. For each simulation type we carry out runs for a cosmology with a sum of neutrino masses
$\sum m_\nu = 0.2$~eV (more details on the cosmology are given in the next section), with an ensemble of $2048^3$ particles for CDM and baryons
and about $21$ billion neutrino particles. In order to study convergence on small and large scales we use two different box sizes, $2$~Gpc/$h$
and $4$~Gpc/$h$, giving us a total of six different runs for this study. We also use the same random number sequence to set up the
perturbations in each type of simulation in order to suppress effects of realization scatter.

As the Newtonian results should be interpreted in N-body gauge, we choose this gauge for our comparison. The runs performed with the
general relativistic approach use Poisson gauge in the simulation, however, at the time the output is written we apply the appropriate
active coordinate transformation to convert the results into N-body gauge. This is done exactly in the same way as it was done for the
initial data of the Newtonian runs, only that this time it is done at the end of the run and not at the beginning. Since the transformation
is computed within linear theory one may be worried that there could be a problem at nonlinear scales; however, the coordinate change goes
to zero on those scales, simply because the Poisson gauge assumes the Newtonian limit discussed at the beginning of this section.

\begin{figure}[t]
\begin{center}
\includegraphics[width=\textwidth]{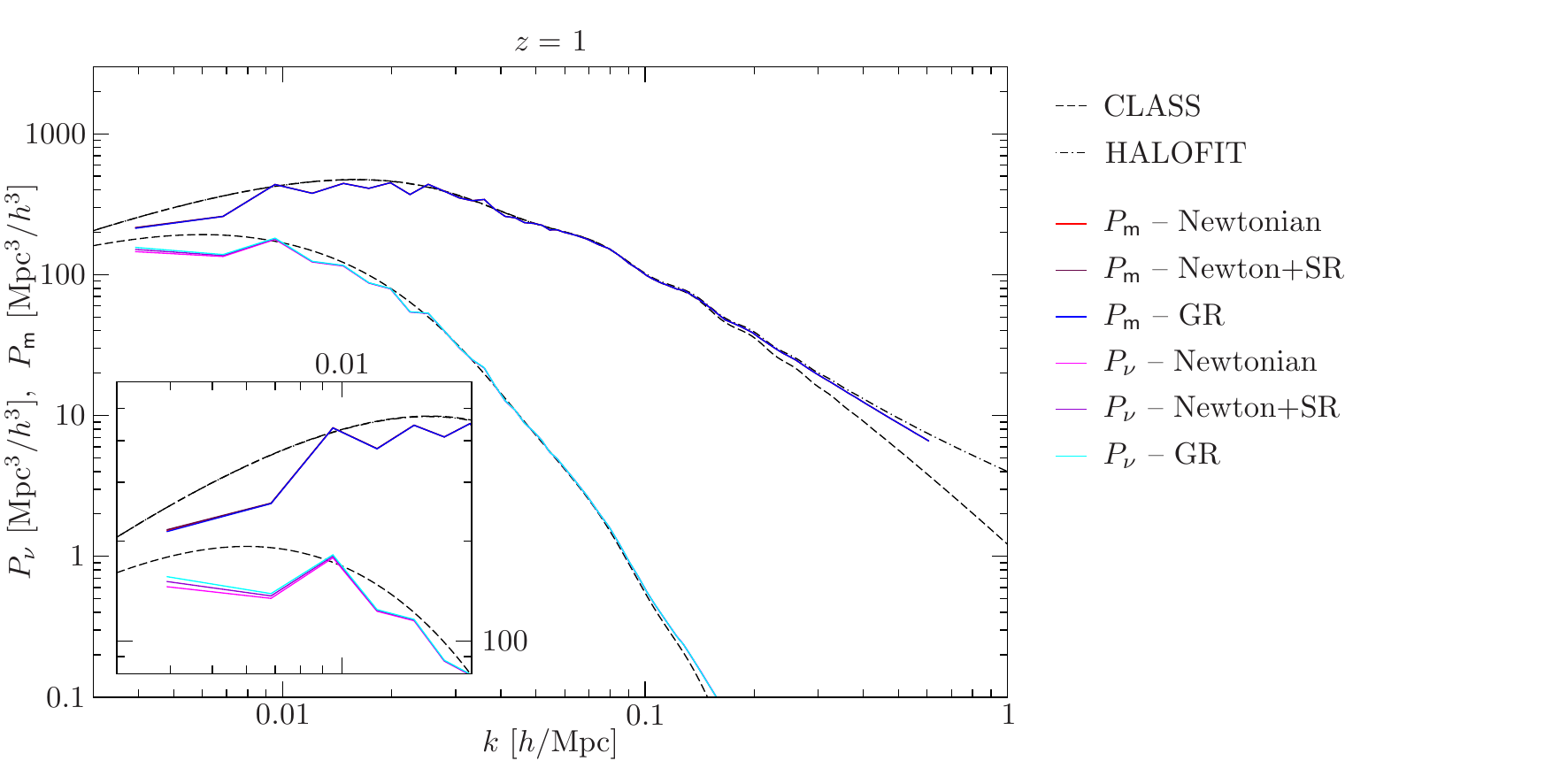}
\end{center}
\caption{\label{f:NvsGR} We show the total matter power spectrum (top curves) and neutrino power spectrum (lower curves) in N-body
gauge at redshift $z = 1$ for three different simulation modes, together with the linear predictions from \textit{CLASS} and the nonlinear
HALOFIT model. The sum of neutrino masses was chosen as $\sum m_\nu = 0.2$~eV for this study. We compare results from a
Newtonian run, a run with Newtonian gravity and special relativity (SR), and a run using our general relativistic (GR)
framework (see text for details).
The inset magnifies the large-scale portion of the spectra where differences in the three simulation modes start to show up.
All runs used the same initial conditions, with a simulation box of $2$ Gpc/$h$, $2048^3$ particles for CDM and baryons, and about $2.1 \times 10^{10}$ neutrino particles.}
\end{figure}

Despite the various approximations it turns out that the results agree remarkably well between the three methods. As shown in
Figure~\ref{f:NvsGR} the nonlinear total matter and neutrino power spectra are practically indistinguishable at small scales. At very large scales
a small disagreement (somewhat less than one per cent on the total matter power at $k \sim 0.005~h/$Mpc) appears whose amplitude seems compatible with the expected error
induced by neglecting neutrino anisotropic stress in the Newtonian simulations \cite{Adamek:2017grt}. As we will show in Section~\ref{sec:relspec}
this effect is fully taken into account in our relativistic runs. The same is true also for the kinetic energy of the neutrinos which may
also give rise to some corrections at high redshift. A possible way to deal with both effects would be to use the ``Newtonian motion gauge''
framework introduced in \cite{Fidler:2016tir}, but for the case of massive neutrinos this still requires some development. In addition, we
expect a small error on the effective free-streaming length of neutrinos when using Newtonian mechanics, which could explain some of the effect we
see in the neutrino power spectrum at large scales. In the end it is hard to disentangle all these contributions in our results.

One other aspect is worth pointing out. When we integrate the particle equations (see Appendix~\ref{app:integrator}) we choose the time
step by requiring that the particles are allowed to move only a certain maximum distance in each leapfrog step. For CDM particles the global
time step (used in the gravity solvers) ensures that this maximum distance is below one grid unit, but for the neutrino particles we may have
to choose smaller time steps -- in fact, we even allow some portion in the tail of the distribution to travel by more than one grid unit.
Applying the same maximum distance criterion in all simulations, the ones with Newtonian mechanics end up needing many more time steps
for integrating the neutrino equations due to the aforementioned issue of superluminal velocities. These simulations therefore consume
significantly more computational resources than the ones with special relativity, even though equations (\ref{eq:SRgeodesic}) and
(\ref{eq:SRvelocity}) contain additional operations. We therefore conclude that the use of the special relativistic equations not only
restores causality but also may prove more economic for simulations that employ a time step criterion based on particle velocities.
However, the common practice so far has been to use a time step criterion that is independent of the neutrino velocity,
thus allowing the neutrinos to move by an arbitrary distance in each step.

\section{Description of numerical simulations}
\label{sec:sims}

\begin{table}[tb]
\begin{centering}
\begin{tabular}{c c c c c c c c}
 $L_\mathsf{box}$ & resolution & $\!\!N_\mathsf{cdm+b}\!\!$ & $N_\nu$ & $m_1$ & $m_2$ & $m_3$ & $\sum m_\nu$ \\ \hline
 $2$ Gpc/$h$ & $0.5$ Mpc/$h$ & $4096^3$ & --- & $0$ eV & $0$ eV & $0$ eV & $0$ eV\\
 $2$ Gpc/$h$ & $0.5$ Mpc/$h$ & $4096^3$ & $1.7\times10^{11}$ & $0$ eV & $0.009$ eV & $0.051$ eV & $0.06$ eV\\
 $2$ Gpc/$h$ & $0.5$ Mpc/$h$ & $4096^3$ & $1.7\times10^{11}$ & $0.0225$ eV & $0.0225$ eV & $0.055$ eV & $0.1$ eV\\
 $2$ Gpc/$h$ & $0.5$ Mpc/$h$ & $4096^3$ & $1.7\times10^{11}$ & $0.05$ eV & $0.05$ eV & $0$ eV & $0.1$ eV\\
 $2$ Gpc/$h$ & $0.5$ Mpc/$h$ & $4096^3$ & $1.7\times10^{11}$ & $0.06$ eV & $0.06$ eV & $0.08$ eV & $0.2$ eV\\
 $2$ Gpc/$h$ & $0.5$ Mpc/$h$ & $4096^3$ & $1.7\times10^{11}$ & $0.1$ eV & $0.1$ eV & $0.1$ eV & $0.3$ eV\\
 $0.9$ Gpc/$h$ & $0.25$ Mpc/$h$ & $3600^3$ & --- & $0$ eV & $0$ eV & $0$ eV & $0$ eV\\
 $0.9$ Gpc/$h$ & $0.25$ Mpc/$h$ & $3600^3$ & $1.2\times10^{11}$ & $0.06$ eV & $0.06$ eV & $0.08$ eV & $0.2$ eV\\
 $8$ Gpc/$h$ & $2.0$ Mpc/$h$ & $4096^3$ & --- & $0$ eV & $0$ eV & $0$ eV & $0$ eV\\
 $8$ Gpc/$h$ & $2.0$ Mpc/$h$ & $4096^3$ & $1.7\times10^{11}$ & $0.06$ eV & $0.06$ eV & $0.08$ eV & $0.2$ eV\\
\end{tabular}
\end{centering}
 \caption{\label{t:sims} Main characteristics of our production runs. These simulations used a total of $2.2$ million
 CPU-hours on the Cray XC50 \textit{Piz Daint} at the Swiss National Supercomputing Centre.}
\end{table}

For the remainder of the paper we discuss results from a large suite of N-body simulations carried out with \textit{gevolution}.
This is the first time that such a comprehensive numerical study is based entirely on a relativistic approach. As summarized in Table~\ref{t:sims}
we explore the mass range between $\sum m_\nu =$ 0.3 eV and the minimal mass scenario, $\sum m_\nu =$ 0.06 eV, and also include a case
where we can compare a normal mass hierarchy ($m_1 \lesssim m_2 < m_3$) to an inverted one ($m_3 < m_1 \lesssim m_2$) at fixed total
mass $\sum m_\nu =$ 0.1 eV. Except for the minimal mass scenario we neglect the smaller $\Delta m_{21}^2$ mass splitting, therefore
running with no more than two distinct mass eigenstates. We choose a relatively large
simulation volume with a box size of $L_\mathsf{box} =$ 2 Gpc/$h$ in order to have good statistics at nonlinear scales. This also
ensures that two important scales are contained in our dynamical range: the equality scale at $\sim$0.4~Gpc/$h$ which characterizes
the peak of the matter power spectrum, and the neutrino free-streaming length which (depending on mass) is typically around 1.5 Gpc/$h$.
These simulations have a spatial resolution of $0.5$ Mpc/$h$ and contain $4096^3$ particles for the CDM and baryonic component,
corresponding to a mass resolution of roughly $10^{10} M_\odot/h$.

The neutrino components are simulated using N-body ensembles with a total of $1.7\times10^{11}$ tracers which are evolved as if they
were fundamental particles of the species they represent. These tracers sample the neutrino phase space and are used to construct a Monte Carlo representation of the
neutrino stress-energy tensor. Since the number of tracers is much smaller than the true number of fundamental particles, the total
stress-energy is obtained by multiplying the one of the tracers by a corresponding large factor. This makes sense since the stress-energy
is an extensive quantity and the ensemble of tracers represents a fair sample. In the cases where several different mass eigenstates are simulated, the size of the sub-ensembles for
the different masses is chosen such that in the nonrelativistic limit each tracer has the same fractional contribution to the stress-energy.
This also means that massless states are not represented as an N-body ensemble -- they are included in the background radiation density
but their perturbations are neglected. This is justified since radiation perturbations decay inside the horizon.

In order to study finite-volume as well as resolution effects, we carry out two pairs of additional production runs. One has a larger
box of $L_\mathsf{box} =$ 8 Gpc/$h$ and hence significantly more volume but lower resolution, the other one has a smaller box of
$L_\mathsf{box} =$ 0.9 Gpc/$h$ with a resolution of $0.25$ Mpc/$h$. At given $L_\mathsf{box}$ all simulations use exactly the same
random number sequence to set up the initial perturbations. Therefore, if we take ratios of numerical power spectra, the cosmic variance
will cancel on linear scales. In order to test various systematics we additionally run over 30 smaller simulations like the ones
discussed in Section \ref{sec:newtonian}.

The following values of cosmological parameters are common to all simulations: $A_s = 2.215\times10^{-9}$ (at $k_\mathsf{pivot} = 0.05$ Mpc$^{-1}$),
$n_s = 0.9619$, $h = 0.67556$, $\omega_\mathsf{b} = 0.022032$, $T_\mathsf{cmb} = 2.7255$ K, $N_\mathsf{eff} = 3.046$ (in the 
ultrarelativistic limit) and no spatial curvature. We want to maintain a fixed total matter density and hence set $\omega_\mathsf{cdm} = 0.12038 - \sum m_\nu /$ (93.14 eV) according
to the neutrino mass contribution. We assume the standard scenario of neutrino freeze-out to set the initial parameters of the neutrino
distribution function.

\begin{figure}[t]
\begin{center}
\includegraphics[width=\textwidth]{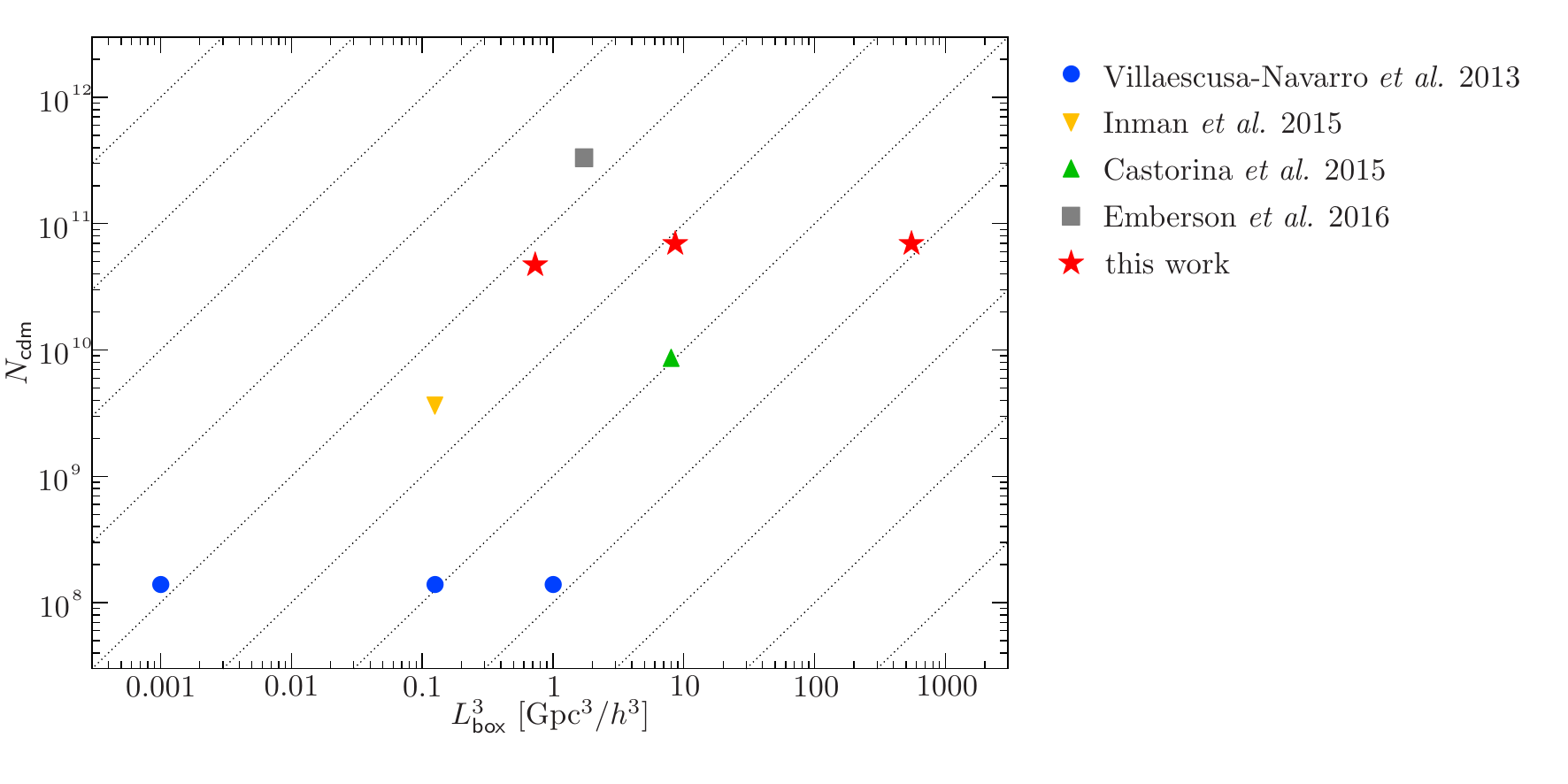}
\end{center}
\caption{\label{f:sims}Diagrammatic comparison of our production runs with some recent N-body simulations
carried out by other groups \cite{VillaescusaNavarro:2012ag,Inman:2015pfa,Castorina:2015bma,Emberson:2016ecv}
to study massive neutrinos. The dotted lines show the lines of constant mass resolution in the CDM component.
The number of neutrino particles per CDM particle varies between $N_\nu/N_\mathsf{cdm} =$ 1 and $N_\nu/N_\mathsf{cdm} =$ 8; we run at
$N_\nu/N_\mathsf{cdm} =$ 2.5.}
\end{figure}

Figure \ref{f:sims} shows a diagram comparing large N-body simulations with massive neutrinos carried out over the past five years.
The studies \cite{VillaescusaNavarro:2012ag,Castorina:2015bma} use a modified version of the code \textit{Gadget}
\cite{Springel:2005mi}, whereas \cite{Inman:2015pfa,Emberson:2016ecv} use the code \textit{CUBEP$^\mathit{\,3\!}$M} 
\cite{HarnoisDeraps:2012vd}. Both are Newtonian N-body codes based on hybrid implementations that combine a particle-mesh
approach on large scales with a direct particle-particle interaction on small scales. Our code currently lacks the latter
feature and is hence less suited for analyzing the detailed evolution of small structures. On the other hand, for the first time
we are able to perform a self-consistent ab initio simulation including all relativistic effects. The regular lattice used for the
particle-mesh algorithm in \textit{gevolution} also allows for a very efficient parallelization that can be scaled up to very
large problem sizes. A typical production run with massive neutrinos (see Table~\ref{t:sims}) parallelized over 16384 CPUs of the
Cray XC50 supercomputer \textit{Piz Daint} consumed about 270\,000 CPU-hours, approximately corresponding to 16 hours wall clock time.
As the diagram in Figure~\ref{f:sims} shows, we present in this paper some of the largest neutrino N-body simulations performed to date in terms of
particle number, second only to \cite{Emberson:2016ecv} which is however just one single simulation -- all other studies ran a suite of
simulations to explore various masses.

\section{Total matter and neutrino power spectra}
\label{sec:matterspec}

\begin{figure}[t]
\begin{center}
\includegraphics[width=\textwidth]{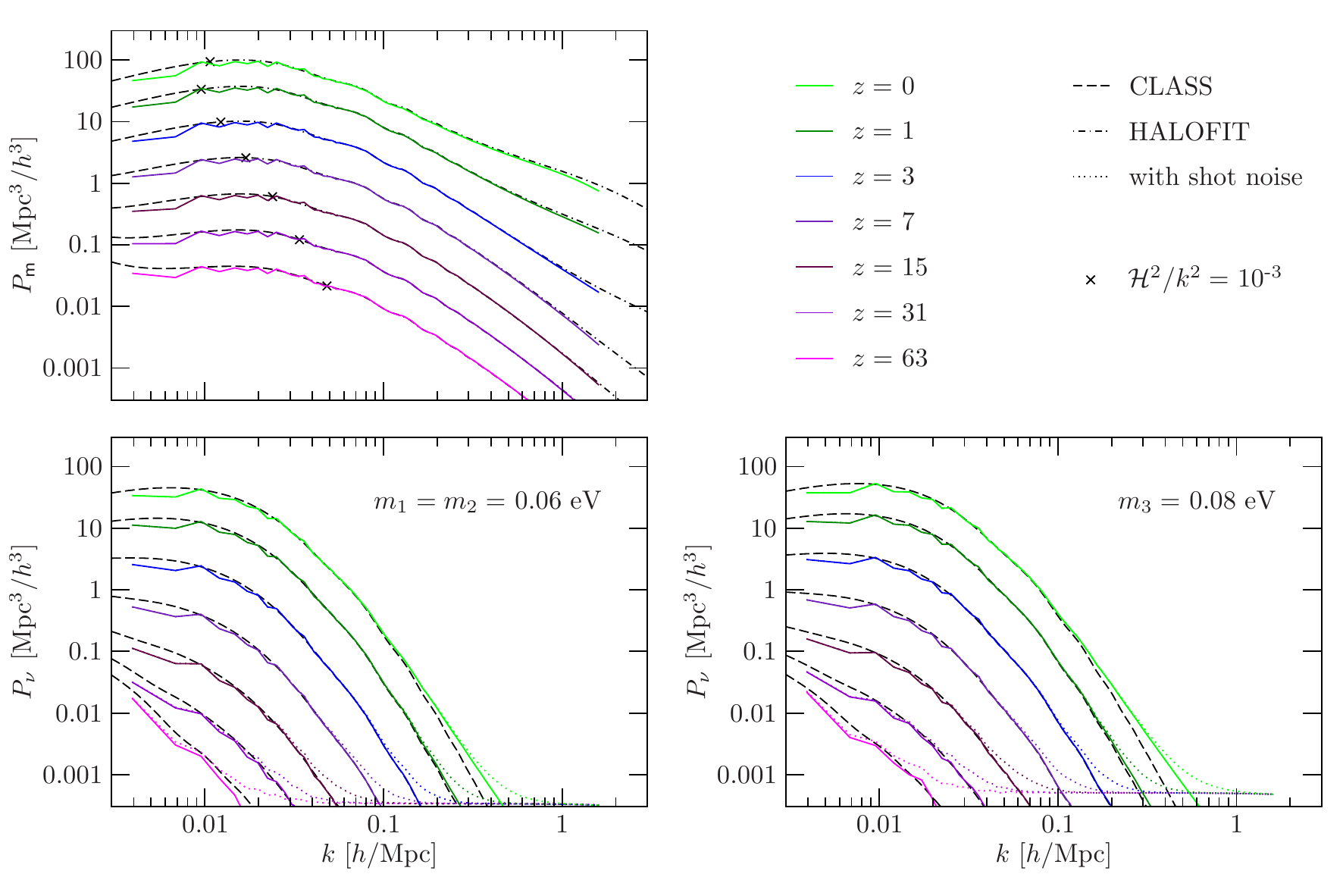}
\end{center}
\caption{\label{f:Pm200} For the case $\sum m_\nu = 0.2$ we show the numerical power spectra in Poisson gauge of total matter,
$P_\mathsf{m}$, as well as those of the individual neutrino components, $P_\nu$, evolving over redshifts from $z =$ 63 to $z =$ 0. The
numerical data is from our simulation with $L_\mathsf{box} =$ 2 Gpc/$h$. For the neutrino components we also show the spectra including the
shot noise (dotted lines) in order to highlight the low noise level we can achieve, characterized by the amplitude at which these spectra level off
horizontally at the bottom of each panel. The dashed black lines indicate the corresponding linear spectra obtained from \textit{CLASS},
whereas the dash-dotted lines show the HALOFIT model of $P_\mathsf{m}$ for nonlinear scales. Since the HALOFIT power spectrum matches to
synchronous gauge at linear scales, we only use it deep inside the horizon where gauge effects $\propto \mathcal{H}^2/k^2$ are strongly
suppressed. For each curve we mark the point at which $\mathcal{H}^2/k^2$ reaches 10$^{-3}$, and at larger scales we again show the linear
spectrum in Poisson gauge instead.}
\end{figure}

Figure \ref{f:Pm200} shows the total matter power spectra (top panel) and neutrino power spectra (bottom panels) for a series of
redshifts for the case $\sum m_\nu =$ 0.2 eV. We adopt the usual definition that total matter consists of CDM, baryons and massive
neutrinos. The density contrast was computed in Poisson gauge, as this is the gauge that underpins
our relativistic framework. Indeed one may notice the characteristic change of slope at scales outside the horizon. As is evident from
the neutrino power spectra, sampling the neutrino component with more than 10$^{11}$ particles allows us to obtain a very low level of
shot noise and thus to prevent any spurious clustering even at high redshift. Especially at low redshift the signal-to-noise
ratio is good enough to see nonlinear effects in the neutrino component alone. At large scales all spectra show good agreement with
linear theory (as computed with \textit{CLASS}). In the late Universe nonlinear effects appear at smaller scales.

The shot noise in the neutrino power spectra is due to the auto-correlation of self-pairs and there are different possibilities
to eliminate or subtract it \cite{Tegmark:1997yq}. We choose to split each neutrino N-body ensemble randomly into two sub-ensembles and 
use their cross-correlation as an estimator for the power spectrum like in \cite{Inman:2015pfa}. This completely eliminates the 
contribution of the self-pairs. We will later use the same method to suppress shot noise in the power spectrum of $\chi$ which 
linearly depends on the neutrino anisotropic stress, see \cite{Adamek:2016zes} for more details.

We want to compare our numerical results to the HALOFIT model, a nonlinear recipe commonly used in the literature to model the
matter power spectrum. We use the implementation provided by \textit{CLASS} which is based on \cite{Takahashi:2012em} with improvements
from \cite{Bird:2011rb} to include the effects of neutrino masses. It is often claimed that Newtonian codes compute the density contrast
of synchronous gauge. This belief is based on the fact that the Newtonian density contrast agrees with the one of a synchronous gauge
in the linear regime, but unfortunately this is no longer true in the nonlinear case where even the very definition of a synchronous
coordinate system faces challenges. In fact, also at linear scales it is conceptually more accurate to say that Newtonian simulations
perform computations in the N-body gauge, as discussed in Section \ref{sec:newtonian}. The N-body gauge employs the same time slicing
as the synchronous gauge which is precisely the reason why the density perturbations match between the two gauges (but other quantities
do not, e.g.\ the coordinate velocity). At nonlinear scales
it is less clear which gauge one should associate to a Newtonian simulation, as one would have to extend the concept of the N-body gauge.
However, as can be seen from the Newtonian limit discussed in Section \ref{sec:newtonian}, the difference to the Poisson gauge is
expected to be of order $\mathcal{H}^2/k^2$. In Figure \ref{f:Pm200} we mark the point on each curve where $\mathcal{H}^2/k^2 =$
10$^{-3}$ and hence the large-scale gauge effects give a permille correction. This point moves towards larger scales at late times and
is close to $k \simeq$ 0.01 $h$/Mpc at redshifts $z \lesssim$ 3. We therefore conclude that we can directly compare our results from
Poisson gauge to the HALOFIT model, which was calibrated to Newtonian simulations and hence refers to an unspecified gauge that
is sufficiently close to Poisson gauge on scales $k \gtrsim$ 0.01 $h$/Mpc for $z \lesssim$ 3.

\begin{figure}[t]
\begin{center}
\includegraphics[width=\textwidth]{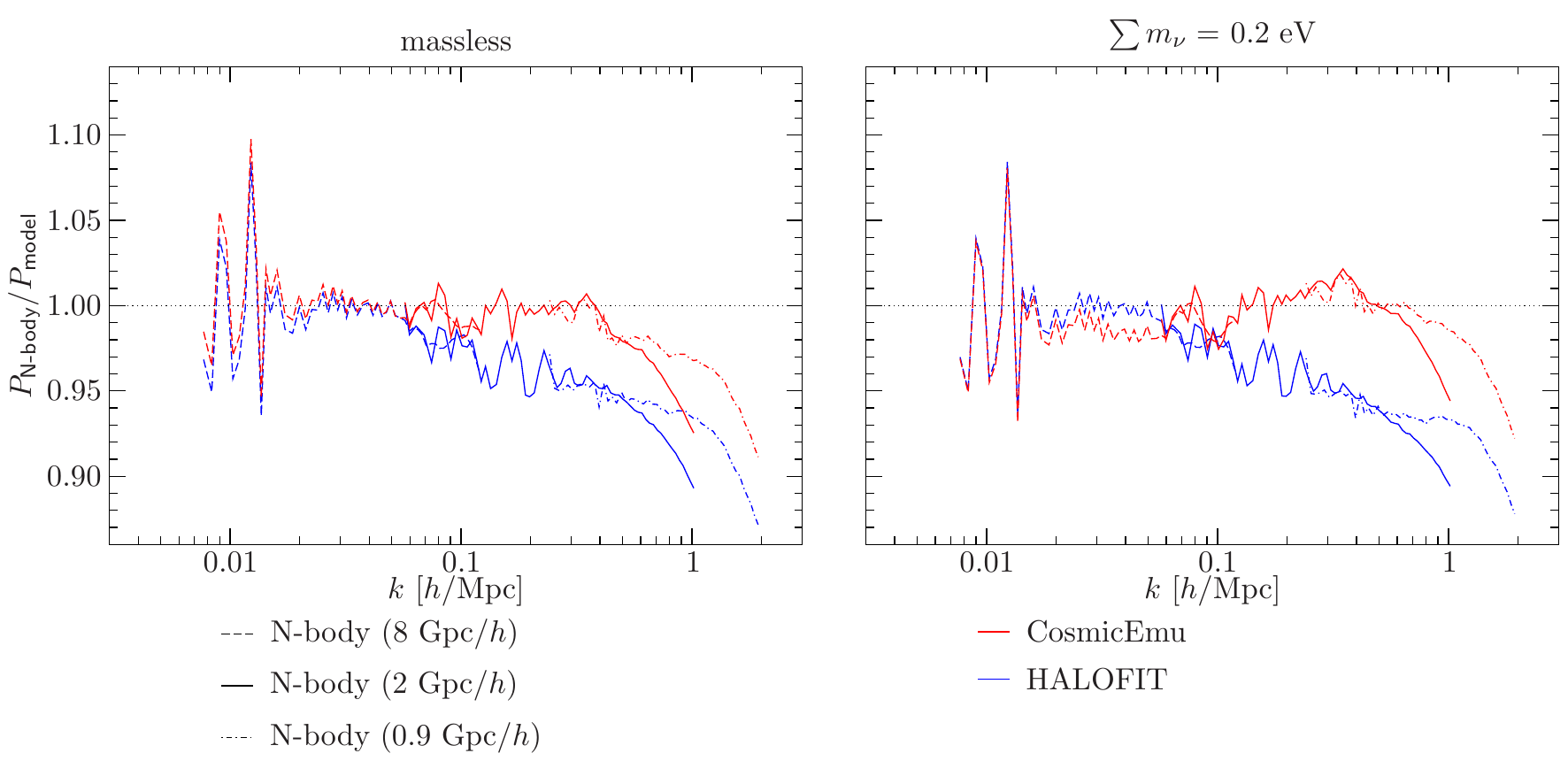}
\end{center}
\caption{\label{f:dPfit}At redshift $z =$ 0 we compare our numerical power spectra of total matter to two nonlinear recipes:
the HALOFIT model (blue) and the latest revision of the Cosmic Emulator (red). We show the massless case and the case
$\sum m_\nu =$ 0.2 eV for which we have three different simulations each to cover a larger range of scales and check for resolution effects.
In both cases the HALOFIT model systematically predicts too much power (the ratio is below unity), about 5\% around the nonlinear scale
$k_\mathsf{nl} \simeq$ 0.3 $h$/Mpc. In the massless case the Cosmic Emulator agrees very well with our results, showing no systematic
deviation at all for $k \lesssim k_\mathsf{nl}$. The agreement is also quite good for the massive neutrino case, although the emulated power spectrum
has a slightly different shape, with more power on large scales and less power on small scales, compared to our numerical result. The
disagreement is however well within the 4\% error budget of the Cosmic Emulator. The comparison between the 2 Gpc/$h$ and the 0.9 Gpc/$h$
boxes shows that the former simulations are converged with respect to resolution effects to about 1\% up to scales $k \lesssim$ 0.6 $h$/Mpc,
and to better than 5\% up to scales of $k \lesssim$ 1 $h$/Mpc.
}
\end{figure}

To get a better impression of the absolute accuracy as well as the robustness with respect to resolution
effects, Figure~\ref{f:dPfit} shows the ratios of our numerical matter power spectra with respect to the HALOFIT model at redshift $z =$ 0.
In addition, we also show the ratios with respect to emulated spectra obtained from the latest revision of the Cosmic Emulator
\cite{Lawrence:2017ost}. While we have a $\sim$ 5\% disagreement with the HALOFIT model, the agreement with the emulated spectra is
excellent, and we therefore suspect that HALOFIT tends to overpredict the total matter power for our chosen cosmology. By comparing
the simulations for different resolutions, i.e.\ the 2 Gpc/$h$ and the 0.9 Gpc/$h$ box, we can convince ourselves that we have a
$\sim$ 1\% numerical convergence of the absolute matter power in the 2 Gpc/$h$ simulations up to $k \simeq$ 0.6 $h$/Mpc, and the
resolution effects reach $\sim$ 5\% only for $k \gtrsim$ 1 $h$/Mpc. This agrees well with our expectation that resolution effects
on the power spectrum scale as $k^2/k_\mathsf{Ny}^2$ at leading order, where $k_\mathsf{Ny} = \pi$ / (0.5 Mpc/$h$) is the Nyqvist
wavenumber corresponding to the resolution of the simulation. It would therefore seem that e.g.\ the $\sim$ 2\% feature at
$k \simeq$ 0.5~$h$/Mpc in the ratio with respect to the emulated spectrum is actually robust. However, this is well within the error budget
of 4\% claimed for the Cosmic Emulator.

It is worth noting that the emulated spectrum seems to have some slight shape issues in the case of massive neutrinos. In particular,
it shows a noticeable excess of power ($P_\mathsf{N\text{-}body}/P_\mathsf{model} <1$) around $k \simeq$ 0.04 $h$/Mpc where HALOFIT always agrees with our numerical results from
the large-volume simulations. This $\sim$ 2\% feature is absent for the case of massless neutrinos.
After discussing with one of the authors of \cite{Lawrence:2017ost} we think this is due to the way
the emulator is matched to linear theory on large scales. As a consequence of this issue the Cosmic Emulator, despite clearly outperforming
HALOFIT in absolute terms, is not well suited when taking ratios of emulated spectra in order to study the relative impact of neutrino
masses. The HALOFIT model may be 5\% off in absolute terms, but the shape of this error seems to be less dependent on the neutrino mass.

\begin{figure}[t]
\begin{center}
\includegraphics[width=\textwidth]{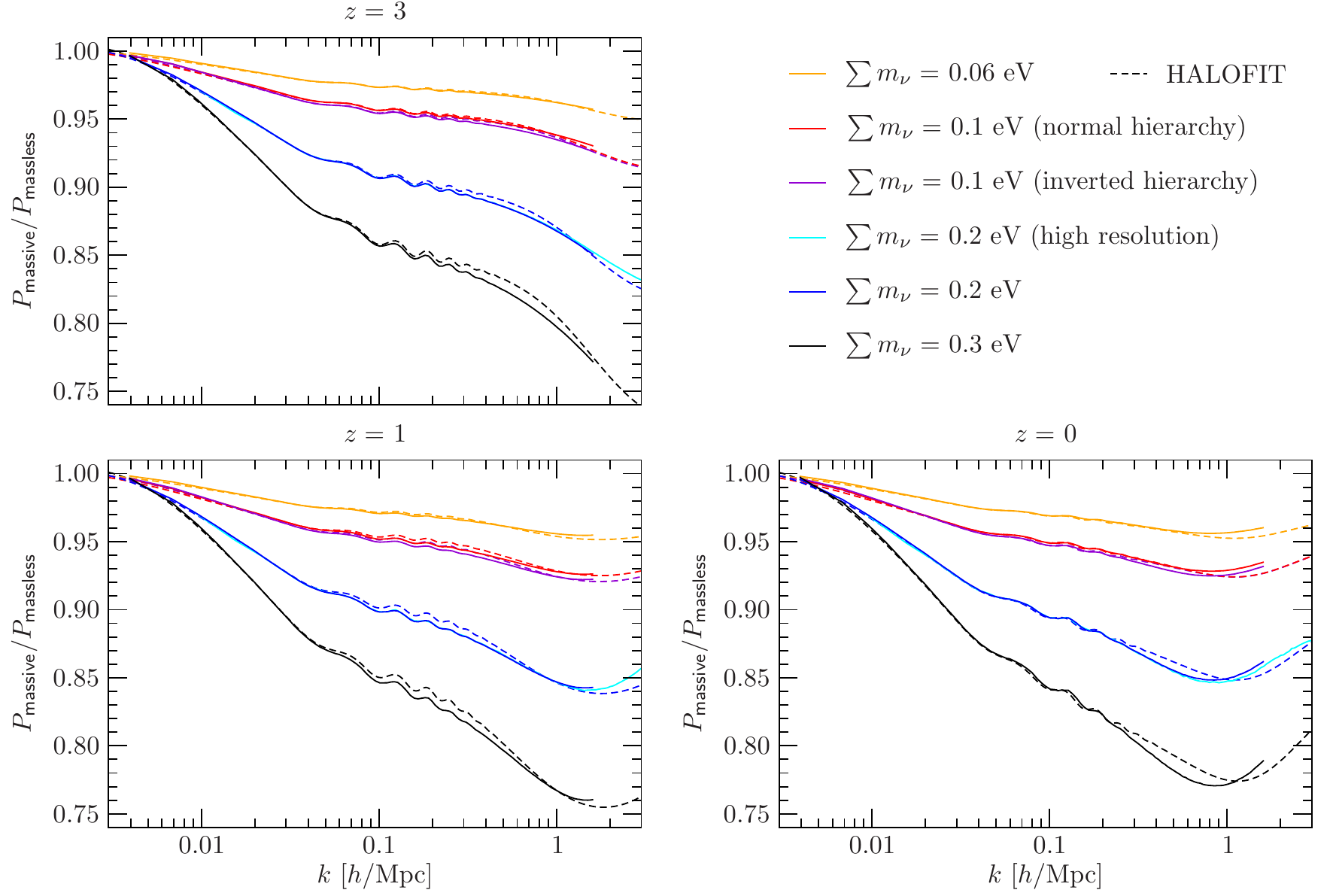}
\end{center}
\caption{\label{f:dPmatter}At three different redshifts we show the ratio of the numerical total matter power obtained for
various neutrino masses relative to the massless case. The numerical data was taken from the runs with $L_\mathsf{box} =$ 2 Gpc/$h$,
except for the light blue curve showing the result for the runs with $L_\mathsf{box} =$ 0.9 Gpc/$h$ that have twice the resolution. The
latter indicates that the relative spectra are converged to better than 1\% for all $k \lesssim$ 1 $h$/Mpc. Dashed lines show the
corresponding predictions from the HALOFIT model.}
\end{figure}

The relative change in the total matter power due to the neutrino masses is shown in Figure~\ref{f:dPmatter} for redshifts
$z =$ 3, 1, and 0. Our resolution study indicates that these results are converged to better than 1\% for all scales $k \lesssim$ 1
$h$/Mpc. This can be understood from the expectation that the leading-order term in the resolution effect will almost cancel when taking
the ratio. To see this, let us schematically write
\begin{equation}
 P_\mathsf{N\text{-}body}(k) \simeq P_\mathsf{continuum}(k) \left(1 + c \frac{k^2}{k_\mathsf{Ny}^2} + \ldots\right)\,
\end{equation}
where $c$ is a coefficient of order unity, and the ellipsis stands for higher-order corrections. If we now take the ratio of two
numerical power spectra one sees that the correction term $\propto k^2/k_\mathsf{Ny}^2$ appears with a coefficient $(c_1 - c_2)$,
where $c_1$, $c_2$ are the respective coefficients for the two spectra in question. It appears reasonable to expect that these do not strongly depend on
cosmology and hence $\vert c_1 - c_2\vert \ll \vert c_1 \vert, \vert c_2 \vert$. In other words, resolution effects are partially
removed when taking ratios of spectra.

Since we keep the total matter density fixed as we vary the neutrino mass, the neutrinos generally lead to a suppression
of the power on scales smaller than their free-streaming length. The suppression is roughly proportional to the sum of the masses.
Intuitively this makes sense since it is the fractional contribution of neutrinos to the total matter which is relevant here. Due
to free streaming the neutrinos constitute a matter component that is much more smoothly distributed than CDM (cf.\ Figure \ref{f:snapshot}). Furthermore, since the
gravitational potential is sourced by total matter, the onset of nonlinear evolution is delayed. This explains why the nonlinear scales
are more sensitive to the neutrino mass. At the other end of the spectrum, at extremely large scales that are outside of the free-streaming
scale, neutrinos effectively behave as CDM and hence a change of their mass has nearly no effect.

Our results show that the relative suppression of matter power is reasonably well modelled by the HALOFIT recipe. The largest disagreement
is seen on mildly nonlinear scales, 0.1 $h$/Mpc $\lesssim k \lesssim$ 1 $h$/Mpc, and can reach $\sim$ 1\%. On these scales HALOFIT generally
underestimates the amount of suppression. Our simulations also indicate that the maximal power suppression is reached at slightly larger
scales than predicted by HALOFIT. Furthermore, we demonstrate that our simulations can distinguish the mass hierarchy at $\sum m_\nu =$ 0.1
eV which is not fully incorporated in the HALOFIT recipe. The hierarchy has an effect of the order of half a per cent on nonlinear
scales, in agreement to what has been found in \cite{Wagner:2012sw}.

\section{Power spectra of relativistic potentials}
\label{sec:relspec}

\begin{figure}[t]
\begin{center}
\includegraphics[width=\textwidth]{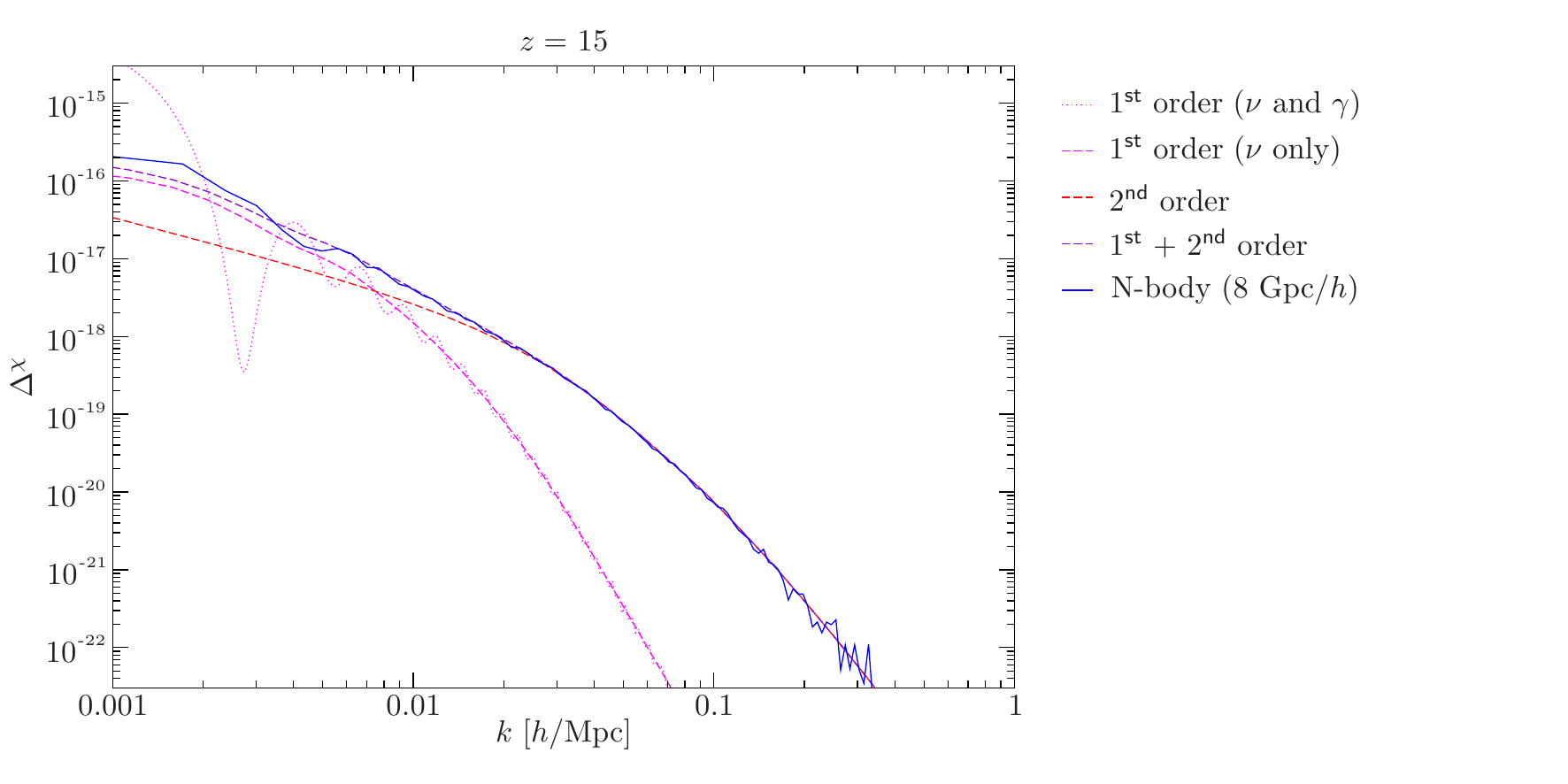}
\end{center}
\caption{\label{f:Pchiz15} At redshift $z = 15$ we show the most important contributions to the power spectrum of $\chi$ for a cosmology 
with $\sum m_\nu = 0.2$~eV. The magenta dotted line is the total first-order contribution from photons and neutrinos (other species do not 
have a first-order contribution) as computed by \textit{CLASS}, which is of interest as we neglect the photon perturbations in
our simulations. The magenta dashed line is the first-order contribution from neutrinos only as computed by \textit{CLASS}. At
second order, also CDM, baryons and geometry (in the guise of a weak-field term $\propto \phi_{,i} \phi_{,j}$) contribute, plotted as
red dashed line. The corresponding expression is given in Appendix~\ref{app:2order}.
The violet dashed line is the total of first and second order, neglecting photons. Our simulation result, shown in blue, is in excellent
agreement with this perturbative prediction. Note in particular that the neutrino N-body ensemble provides the required amount of
anisotropic stress at large scales.}
\end{figure}

In our relativistic N-body code we also calculate the induced vector and tensor perturbations of the metric as well as the
gravitational slip. As we have seen in the previous section, the free-streaming of massive neutrinos gives rise to a suppression
of small-scale power that generally also leads to a suppression of these relativistic effects, as they are mainly sourced by
the nonlinearities in the matter distribution which are strongest on small scales. On the other hand, neutrinos themselves initially have very high velocities which, as we will
discuss below, gives rise to gravitational slip at large scales.

\subsection{Scalar potentials}

In Newtonian gravity the two Bardeen potentials $\phi$ and $\psi$ are equal and they are equal
to the Newtonian potential. In general relativity they are different and we denote their difference by
$\chi = \phi - \psi$.
Within linear perturbation theory $\chi$ is sourced by the scalar anisotropic stress
of relativistic species, i.e.\ photons and relativistic neutrinos. However, this contribution is strongly damped by free-streaming inside
the sound horizon, and after the end of radiation domination the second-order anisotropic stress of CDM starts to dominate on most scales.
Only on very large scales the first-order relativistic contribution remains dominant for a long time. In Figure~\ref{f:Pchiz15}
we show the various contributions at redshift $z=15$ computed in first and second order perturbation theory, together with our
nonperturbative numerical result (shot noise is removed using the technique discussed in Section~\ref{sec:matterspec}). It is worth noting that our neutrino N-body ensemble gives rise to the expected linear effect
at very large scales. We plot the dimensionless power spectra defined by
\begin{equation}
\Delta(k) = 4\pi k^3 P(k) \,.
\end{equation}

As the neutrinos cool down, their anisotropic stress decays and eventually becomes
very subdominant and their main effect then
is the reduction of small scale power and therefore of non-linearities which are the main source of $\chi$ at late times. At late times,
massive neutrinos hence induce a significant suppression to the $\chi$-spectrum, see Figure~\ref{f:Pchi}.
Especially on small scales there is an important suppression of the numerical power-spectrum due to non-linearities compared to the second order result (shown as dashed lines).

\afterpage{
\begin{figure}[p!]
 \includegraphics[width=\textwidth]{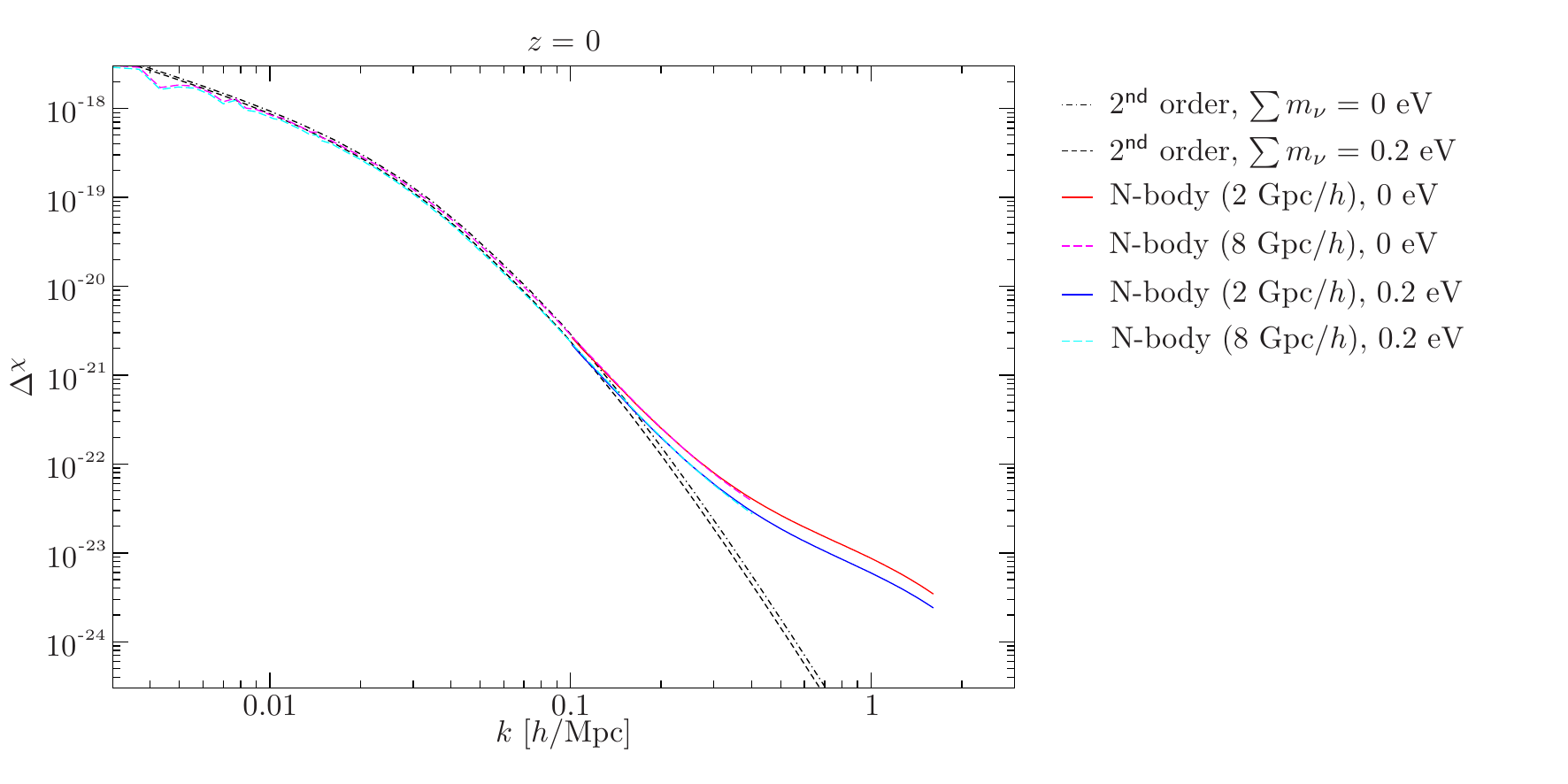}
\caption{\label{f:Pchi} We show the dimensionless power spectra of $\chi$ at redshift $z=0$ for a cosmology with $\sum m_\nu = 0.2$~eV
compared to the case of massless neutrinos. As expected from a second-order calculation (black dashed and dash-dotted lines) the
power is lower in the massive case as a result of the damping of matter perturbations. The first-order anisotropic stress
of the neutrino component has decayed and can now be neglected against the second-order contribution from CDM even at large
sub-horizon scales.}
\end{figure}

\begin{figure}[h!]
 \includegraphics[width=\textwidth]{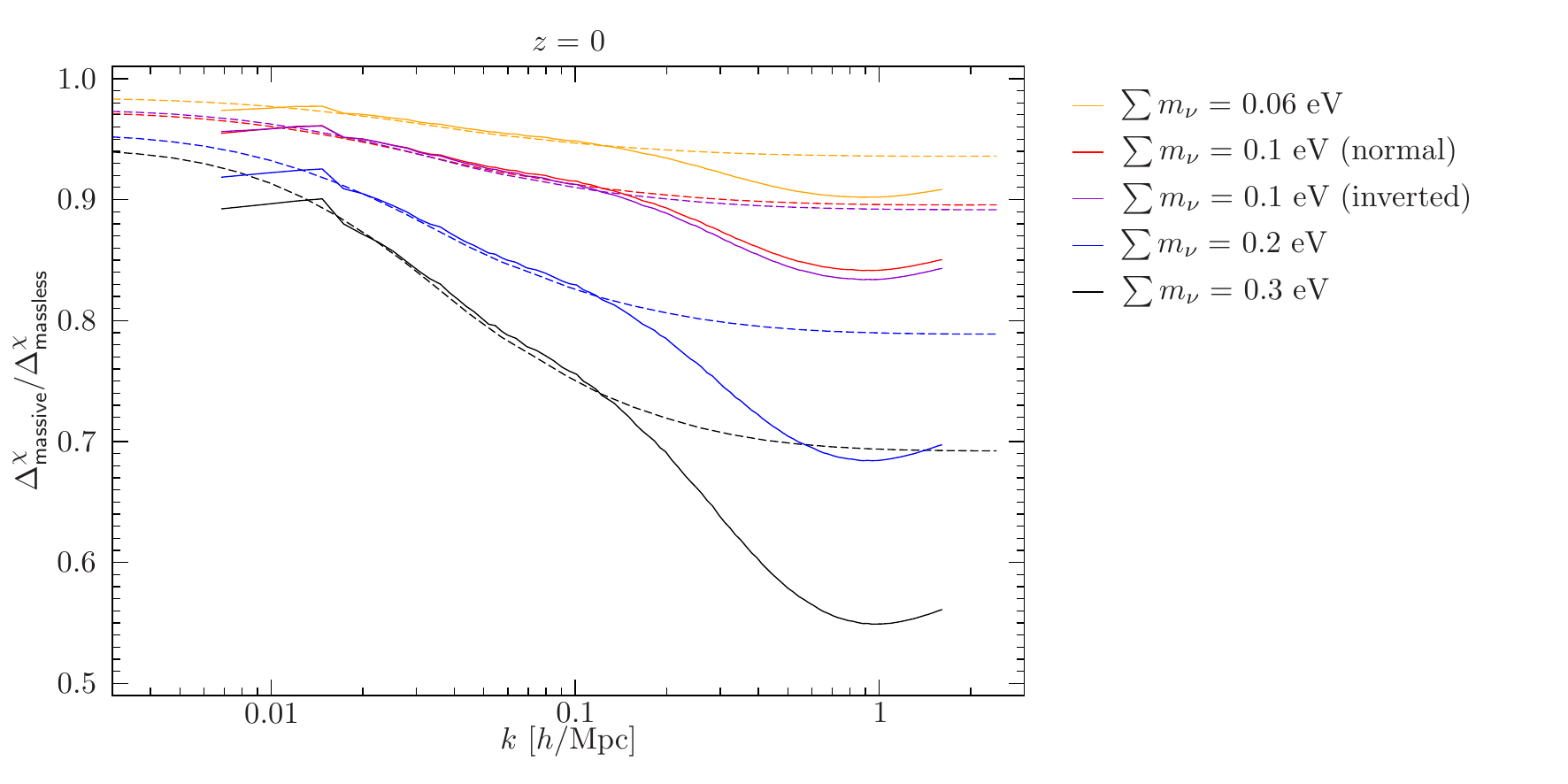}
\caption{\label{f:dPchi} We show the ratios of the $\chi$-spectra for different neutrino masses with respect to the massless case at redshift $z=0$. 
The solid curves are obtained from our relativistic N-body simulations with a 2~Gpc/$h$ comoving box. The dashed lines show the corresponding predictions from a second-order calculation. The onset of non-linear clustering at $k\gtrsim 0.1h$/Mpc is clearly visible. The power suppression in $\chi$ is stronger than in the total matter power spectra, shown in Figure \ref{f:dPmatter}.
}
\end{figure}
\clearpage}

In Figure~\ref{f:dPchi} we compare the ratios
${\Delta^\chi_\mathsf{massive}(k)}/{\Delta^\chi_\mathsf{massless}}(k)$ at $z=0$.
The difference is largest, about 50\%,  for the highest mass, $\sum m_\nu = 0.3$~eV, and smallest, about 10\%, for the smallest mass, $\sum m_\nu = 0.06$~eV.
Note the slight difference between the normal and inverted hierarchy for $\sum m_\nu = 0.1$~eV. We have also found that the relative difference of the spectra increases with time and that its maximum moves slowly to larger scales from $k_{\max}\sim$ 4 $h$/Mpc at redshift $z=3$ to 
$k_{\max}\sim$ 0.7 $h$/Mpc at redshift $z=0$.
 The same behavior is also visible in other spectra. Once a scale has become very non-linear, the
effects of neutrino damping become less important.

\afterpage{
\begin{figure}[t]
 \includegraphics[width=\textwidth]{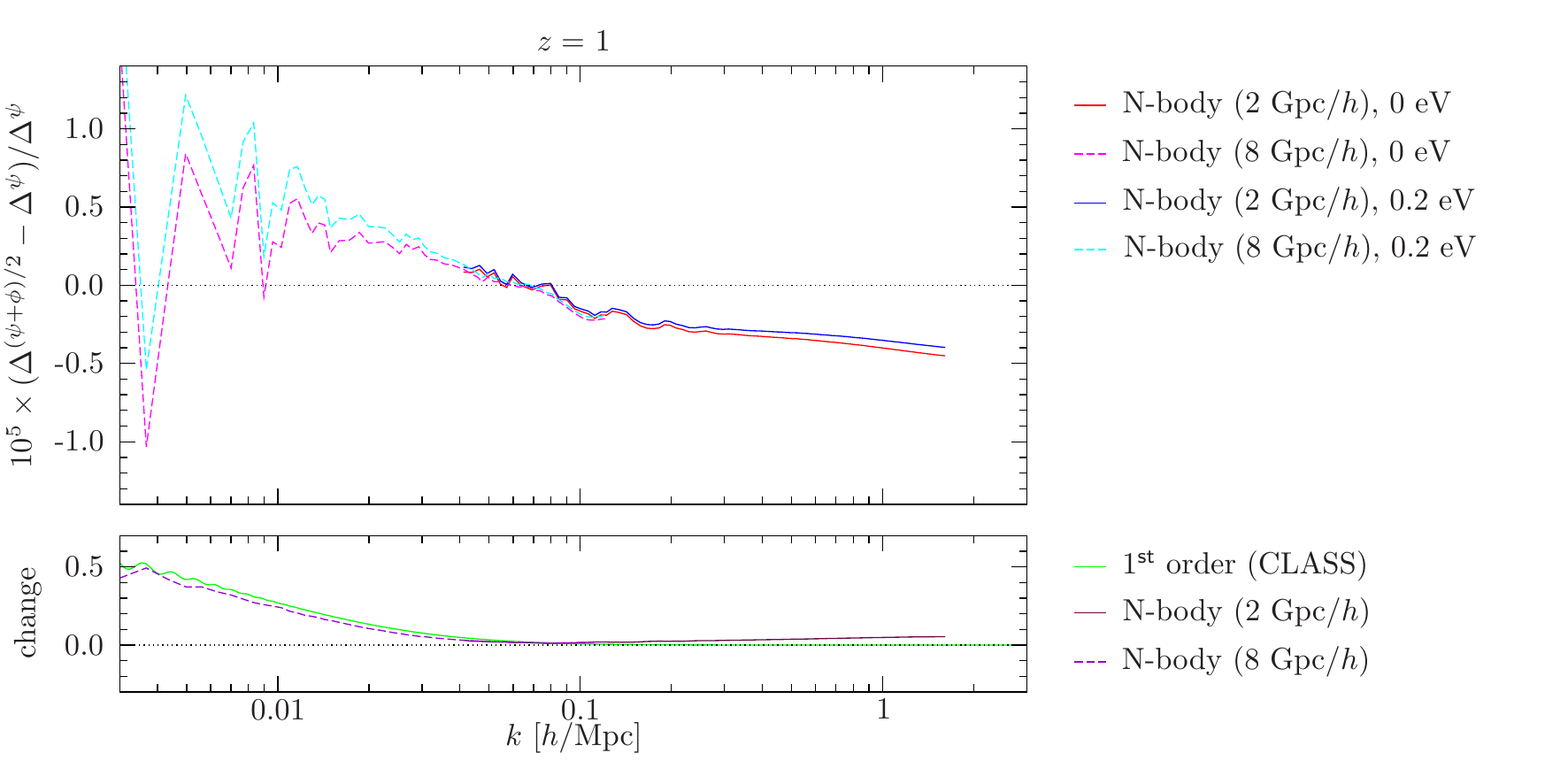} 
 \caption{\label{f:cphichi} We show the relative difference (multiplied by 10$^5$) between the power spectra of the
 Weyl potential $(\psi + \phi)/2$ and the first Bardeen potential $\psi$ for massless neutrinos and for the case
 $\sum m_\nu = 0.2$ eV at redshift $z = 1$ (upper panel). The lower panel shows the change of this quantity between the two cases (in the same units). At large scales, most of the change
 is due to the first-order anisotropic stress in the neutrino component.
 }
\end{figure}
}

In Figure~\ref{f:cphichi} we show the difference between the Weyl potential, $(\phi+\psi)/2$ and the Newtonian potential $\psi$ for
massive and massless neutrinos.  As becomes clear from the lower panel, the change in this difference is quite relevant, about 50\%. On
the other hand, it agrees rather well with the result expected from linear perturbation theory computed with \textit{CLASS} (green 
line). On small scales, when non-linearities are generated, this difference which is purely due to neutrino anisotropic
stresses is less important, and of the order of 5\%. Photons respond to the Weyl potential while massive particles respond to $\psi$;
to see this compare the geodesic equation~(\ref{eq:geodesic}) in the limits $q^2 \gg m^2 a^2$ and  $q^2 \ll m^2 a^2$. Therefore this difference is in principle testable.
This effect is much larger for modified gravity theories but it is important to keep in mind that it is also present in
the standard model, albeit with a very small amplitude of about $10^{-5}$ (see also \cite{Ballesteros:2011cm}).

\subsection{Vector potential}
As we have already found previously~\cite{Adamek:2015eda,Adamek:2016zes}, the 
vector perturbations of the metric are the largest relativistic signature. In amplitude they reach about 1\% of the scalar perturbations
and might therefore in principle be detectable (even though so far a good idea to specifically detect these vector modes is still lacking).

\afterpage{
\begin{figure}[p!]
\begin{center}
\includegraphics[width=\textwidth]{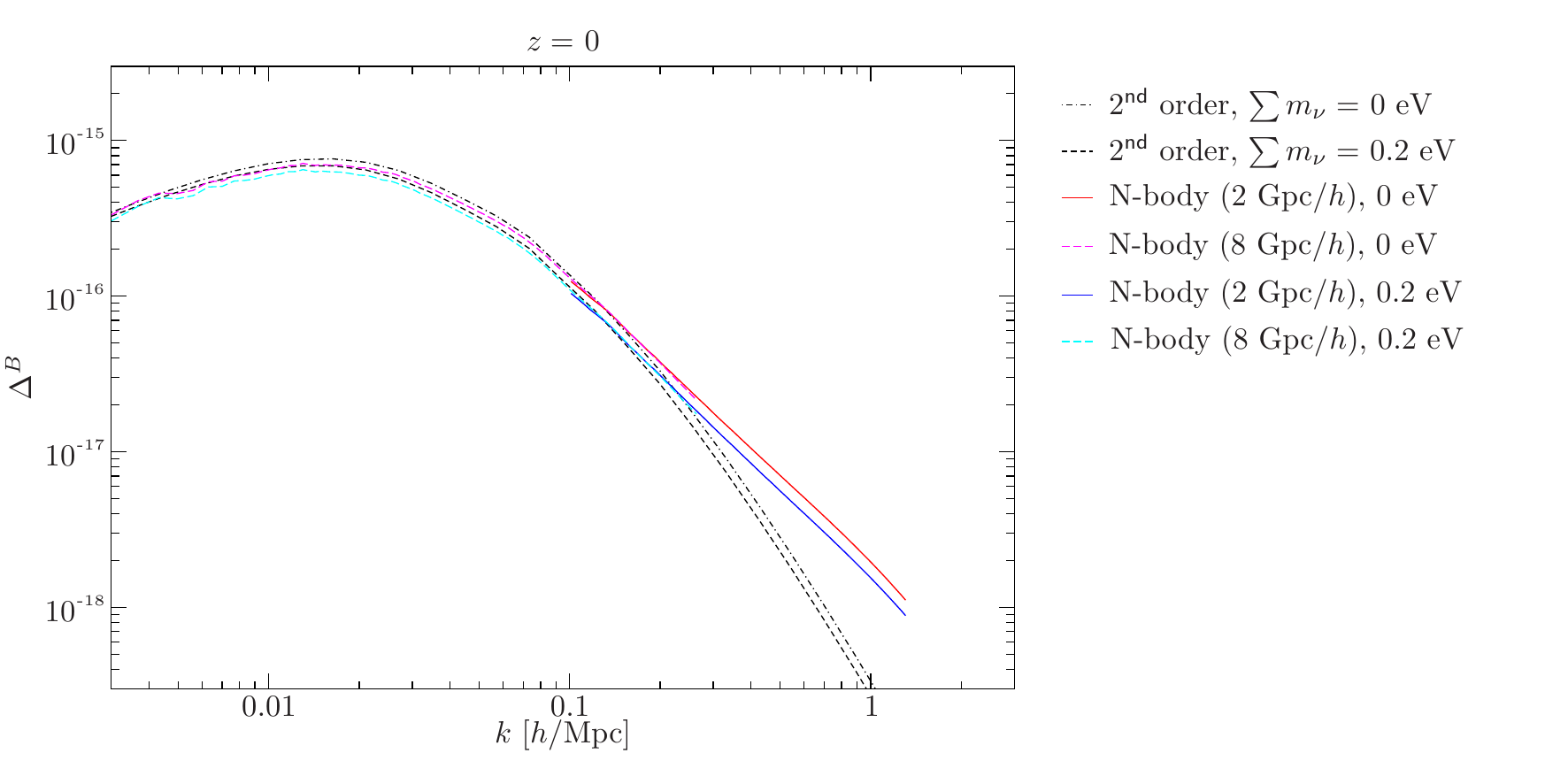}
\end{center}
\caption{\label{f:PB}The vector spectra at redshift $z=0$ for a massless neutrino cosmology and one with $\sum m_\nu = 0.2$~eV. As
in Figure~\ref{f:Pchi} we also show the second-order calculation (black dashed and dash-dotted lines, see Appendix~\ref{app:2order}) indicating the general trend:
the damping of matter perturbations reduces the amplitude of the relativistic potentials.}
\end{figure}

\begin{figure}[h!]
 \includegraphics[width=\textwidth]{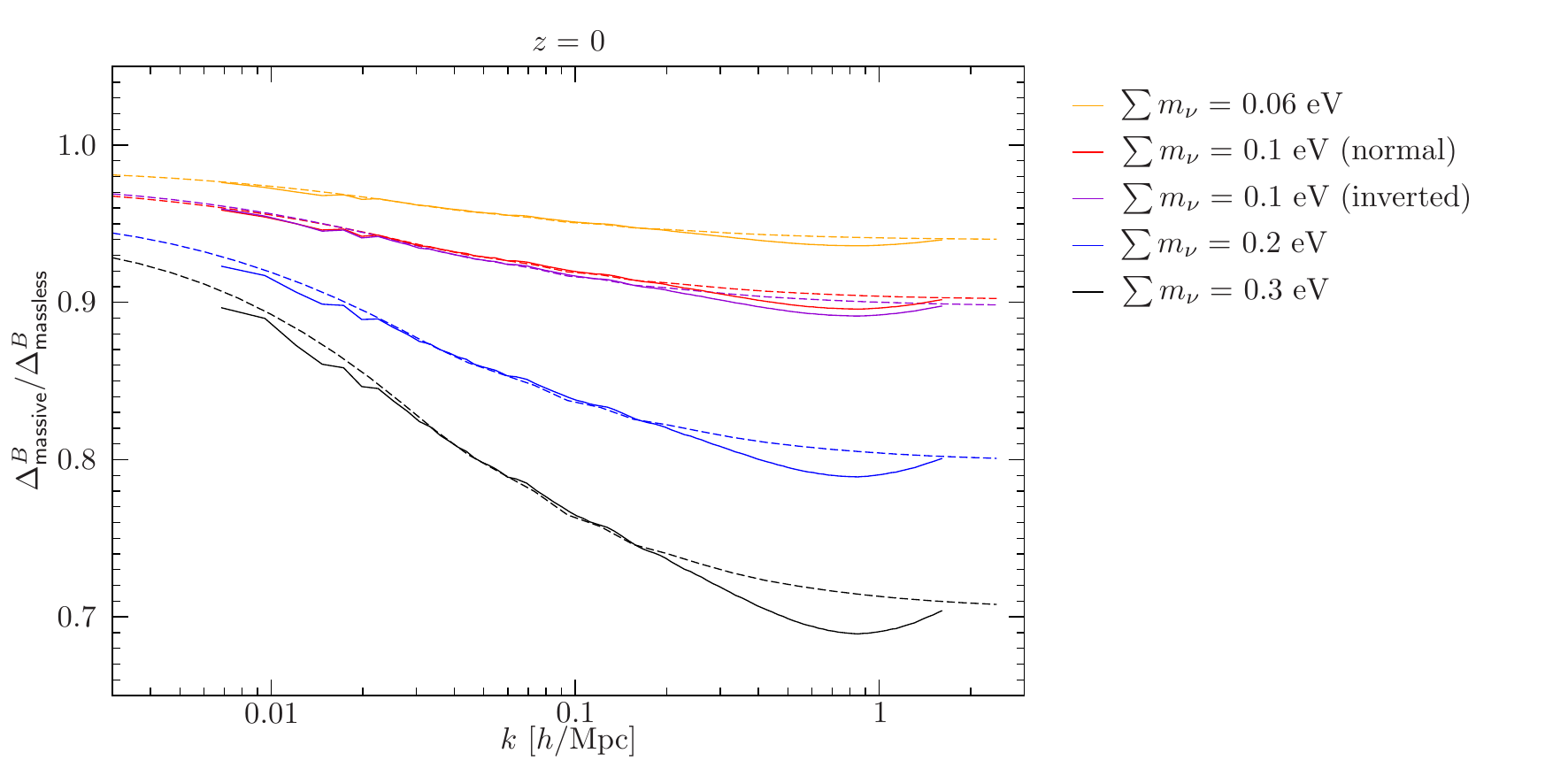} 
\caption{\label{f:dPB}The ratios of the vector spectra for various neutrino masses with respect to the massless case at redshift $z=0$. 
The solid curves are obtained from our relativistic N-body simulations with a 2~Gpc/$h$ comoving box. The dashed lines show the corresponding predictions from a second-order calculation.}
\end{figure}
\clearpage}

In Figure~\ref{f:PB} we show the spectra of vector perturbations at redshift $z=0$  for simulations with $\sum m_\nu =$ 0.2~eV
and with massless neutrinos. The results from a second order calculation are also shown. In Figure~\ref{f:dPB} we show the ratio ${\Delta^B_\mathsf{massive}(k)}/{\Delta^B_\mathsf{massless}(k)}$ at $z=0$ for neutrino masses from $\sum m_\nu =$ 0.3~eV down to
$\sum m_\nu =$ 0.06~eV. Again the difference is largest for the largest neutrino mass. For $\sum m_\nu =$ 0.3~eV the $B$-power spectrum is reduced by about 30\% at the scale $k \simeq 0.7\, h$/Mpc. But also for the smallest
neutrino mass we have considered, $\sum m_\nu =$ 0.06~eV, the reduction still reaches about 5\%.

We have found that the difference in the spectra increases with decreasing redshift on large to intermediate scales. On small
scales, $k\lesssim$ 1 $h$/Mpc the trend is reversed at late time, like for the Newtonian gravitational potential.  On theses small scales, at late times
non-linearities become sufficiently strong so that clustering with neutrinos slowly catches up with the pure CDM case.

\subsection{Tensor perturbations}

In Figure~\ref{f:Ph} we show the tensor power spectrum at redshift $z=0$  for simulations with $\sum m_\nu =$ 0.2~eV and for
massless neutrinos. The second order perturbative result is also indicated as dashed line. In Figure~\ref{f:dPh} we show the ratio
$\Delta^h_\mathsf{massive}(k)/\Delta^h_\mathsf{massless}(k)$ for all masses investigated. Again, the
difference is largest for the largest neutrino mass, $\sum m_\nu =$ 0.3~eV, where it amounts to about 40\% on small scales,
$k \gtrsim$ 0.8 $h$/Mpc. But also for the smallest neutrino mass we have looked at, $\sum m_\nu =$ 0.06~eV, the difference is still
nearly 10\%.
The difference between the normal and the inverted hierarchy, however, does not rise above about 1\% like also for the other spectra.

\afterpage{
\begin{figure}[p!]
\begin{center}
\includegraphics[width=\textwidth]{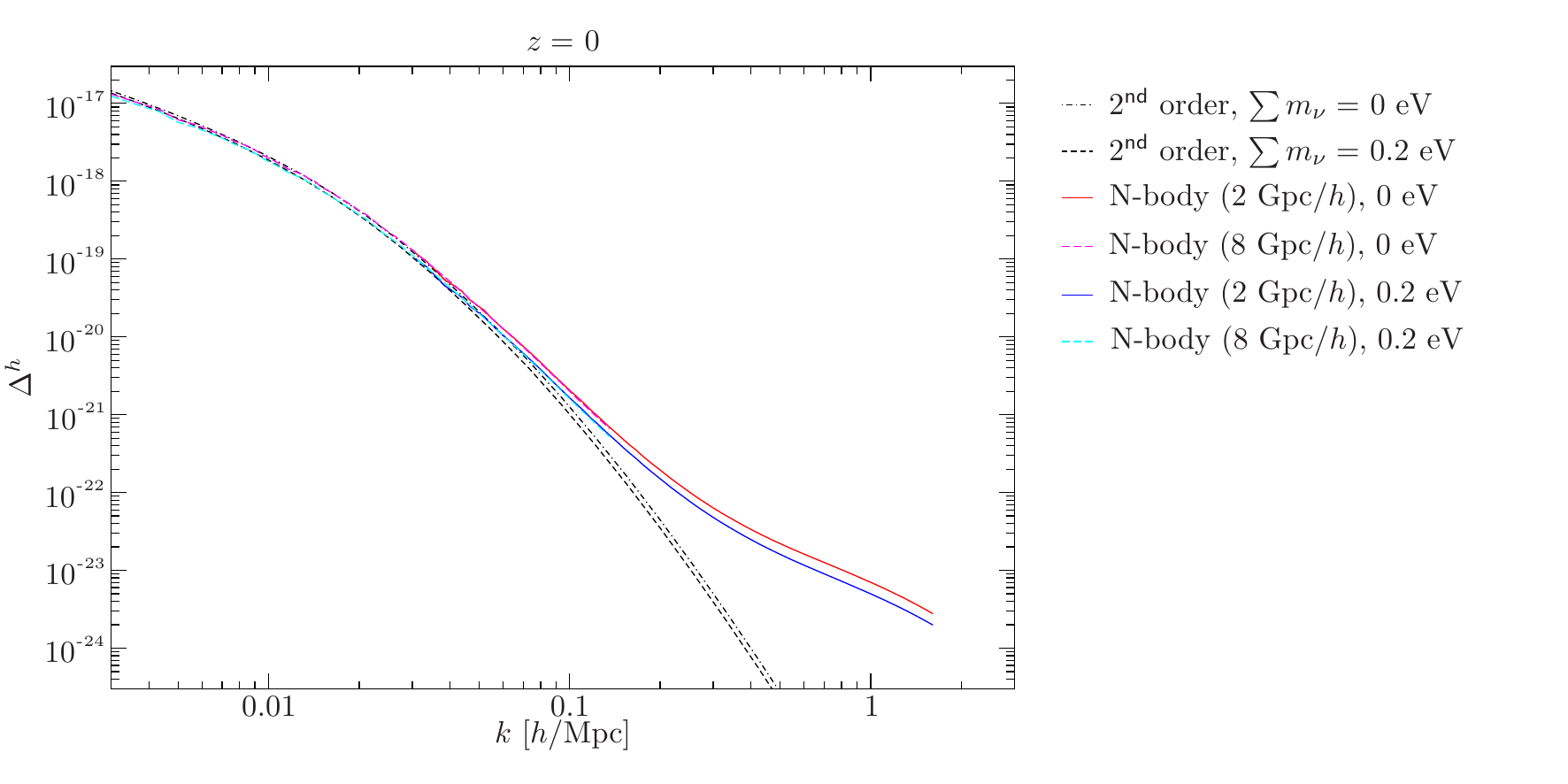}
\end{center}
\caption{\label{f:Ph}The tensor spectra at redshift $z=0$
for a massless neutrino cosmology and for one with $\sum m_\nu = 0.2$~eV. The predictions from the second-order calculation
(see Appendix~\ref{app:2order}) are also shown.
}
\end{figure}

\begin{figure}[h!]
\begin{center}
\includegraphics[width=\textwidth]{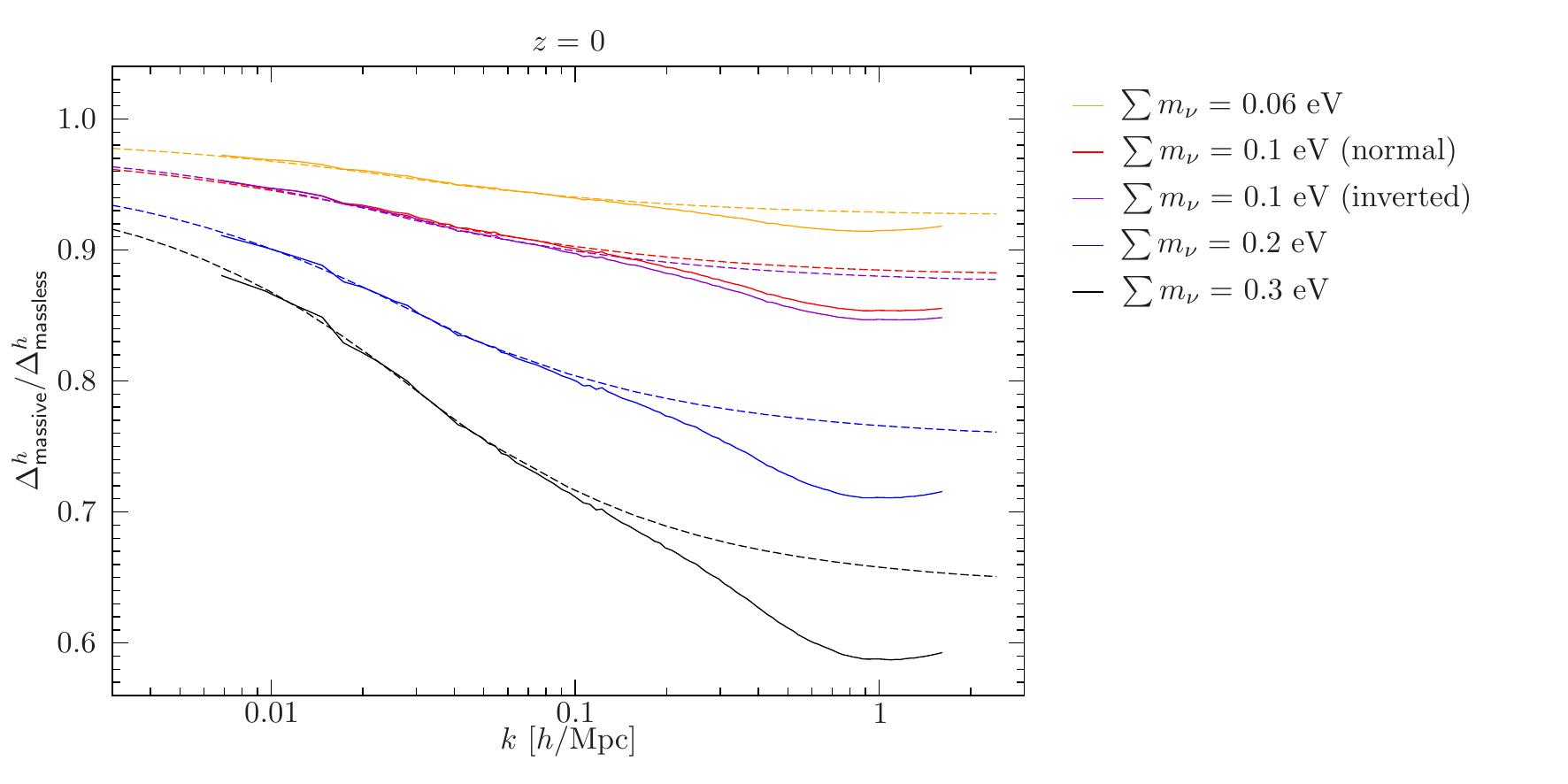}
\end{center}
\caption{\label{f:dPh} We show the ratios of the spectra of the spin-2 metric perturbation for various neutrino masses with respect to the massless case at redshift $z=0$. The solid curves are obtained from our relativistic N-body simulations with a 2~Gpc/$h$ comoving box. The dashed lines show the corresponding predictions from a second-order calculation.
}
\end{figure}
\clearpage}

Finally, we want to stress that these tensor modes are not gravitational waves in the usual sense of the term. They are not freely
propagating but they are actually the adiabatic response of spacetime to the presence of a slowly varying source, the tensor anisotropic stress
of matter. Their time dependence is not wave-like so that one cannot detect these tensor modes in a LIGO-like interferometric experiment.

We have seen in this section that the differences between the relativistic spectra without and with neutrino masses are quite important, up to about
40\% for the tensor spectra, 30\% for the vector spectra, 50\% for the $\chi$-spectra and, correspondingly, for the difference between the Weyl potential and the Newtonian potential. Even for the minimal neutrino masses allowed by the oscillation data, the impact is substantial (up to 5\%-10\%) and persists over a wide range of scales. However,
these spectra are all significantly suppressed with respect to the dominant Newtonian gravitational potential spectrum by at least four orders of
magnitude and none of them has been measured so far at cosmological scales. The most promising is the vector spectrum which is up to 22\%  smaller in the presence
of massive neutrinos with a total mass of $\sum m_\nu =$ 0.2~eV. The amplitude of the vector spectrum is about 1\% of the Newtonian potential and the difference
in the amplitude is about half the difference of the spectrum which would lead to a 10\% effect in the B-mode amplitude or a 0.1\% effect on
the total gravitational field. This is probably not measurable without a new, groundbreaking  idea of how to extract a vector signal.

The relative difference between the Weyl potential and the Newtonian potential is about $10^{-5}$ on large scales. The tensor power spectrum is two orders of magnitude smaller than the vector signal and the $\chi$-spectrum is about 
three orders of magnitude smaller. Their detection is correspondingly even more difficult.

\section{Halo mass function}
\label{sec:hmf}

\begin{figure}[t]
\begin{center}
\includegraphics[width=\textwidth]{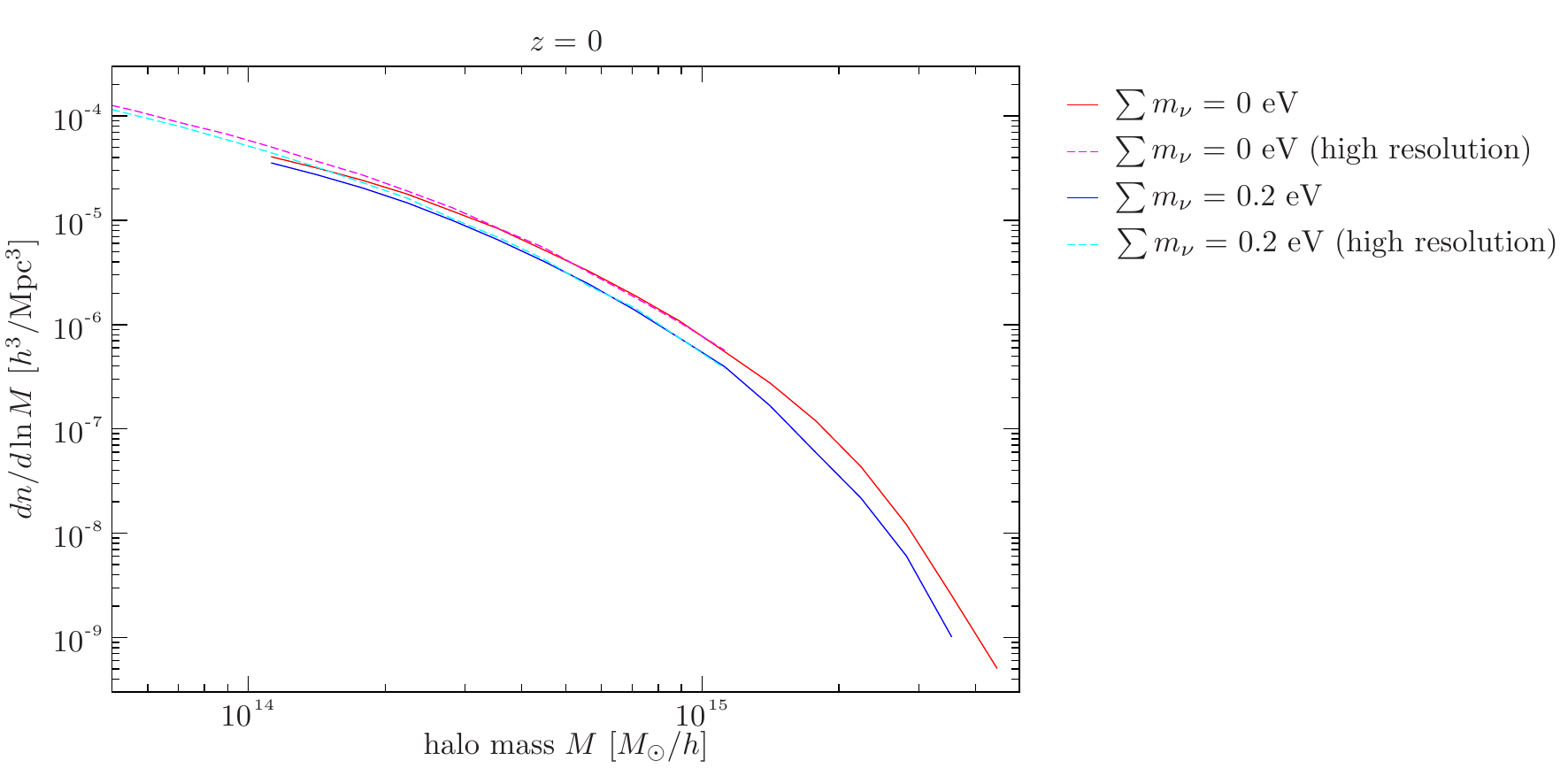}
\end{center}
\caption{\label{f:hmf} We show the halo mass function measured at redshift $z =$ 0 for the case of massless neutrinos and the one
with $\sum m_\nu =$ 0.2~eV. The halo catalogs were obtained from snapshots of the simulations with 2 Gpc/$h$ (solid lines) and 0.9 Gpc/$h$
(dashed lines) boxes. The latter have a higher resolution and hence we can conclude that the numerical results are contaminated by
resolution effects of less than $\sim$10\% for halo masses $M \gtrsim$ 3$\times$10$^{14} M_\odot/h$.}
\end{figure}

Owing to the fact that \textit{gevolution} works at fixed spatial resolution it is not very well suited for studying CDM halos in
great detail. However, the resolution achieved in our production runs is sufficient to obtain robust results for some global 
properties of massive halos, e.g.\ their virial masses and radii. In Figure~\ref{f:hmf} we show the halo mass functions obtained at
redshift $z =$ 0 for the case of massless neutrinos and the one with $\sum m_\nu =$ 0.2~eV. In both cases we use snapshots from simulations
at two different resolutions in order to check for resolution effects. The lack of adaptive force resolution attenuates the formation
of small structures whereas more massive objects will be less affected. Moreover, the virial mass is defined through a spherical
overdensity threshold and will have large errors for objects whose virial radius is not well resolved. As expected, the halo mass
function is therefore most robust at high masses. For those we obtain results that are converged to better than 10\% over a range of
more than one order of magnitude in mass.

\begin{figure}[t]
\begin{center}
\includegraphics[width=\textwidth]{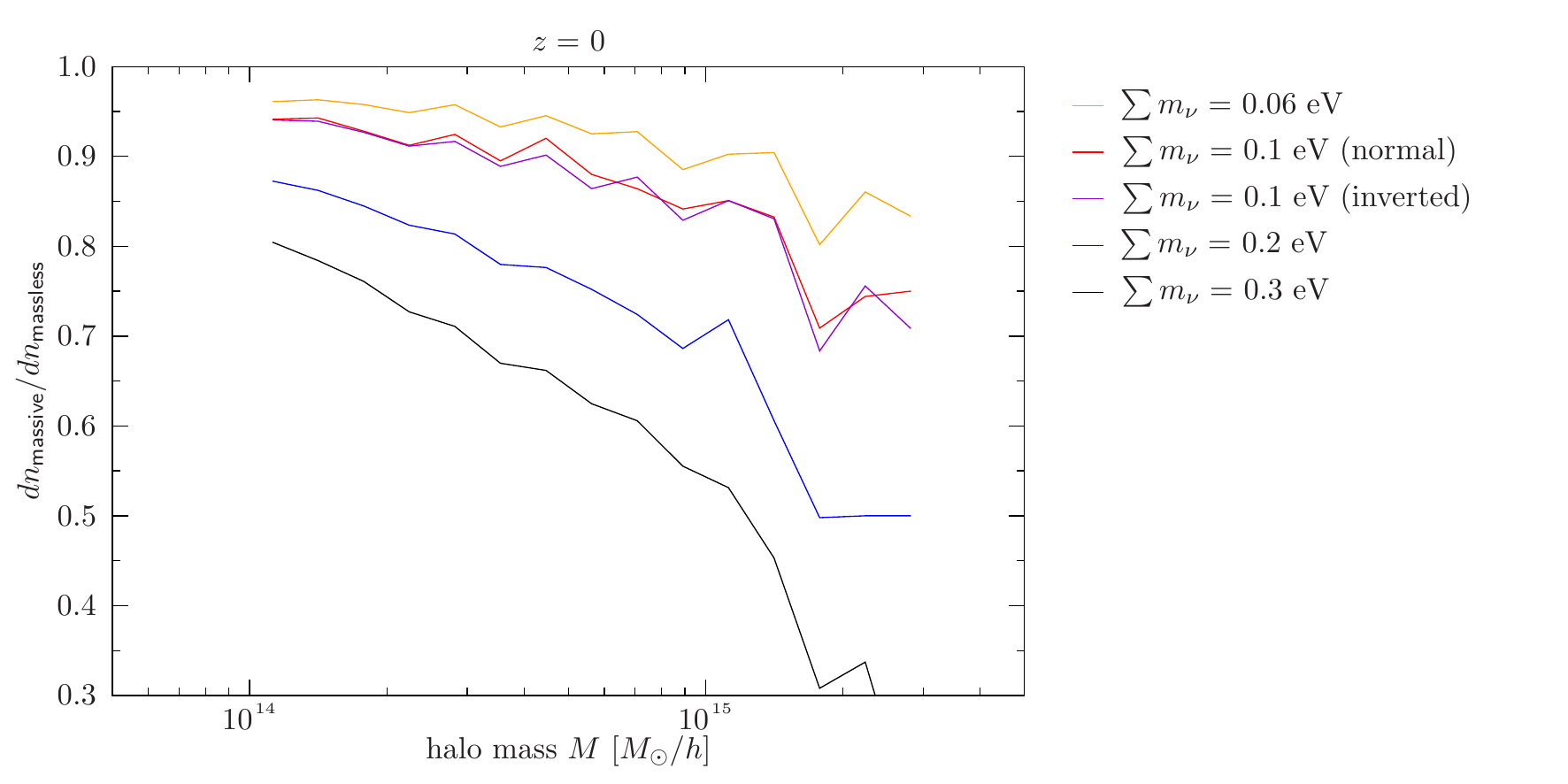}
\end{center}
\caption{\label{f:dhmf} In order to quantify the relative impact of the neutrino mass we show the ratio of the halo mass functions
at redshift $z =$ 0 for different neutrino masses with respect to the massless case. The change is most dramatic for the highest halo
masses.}
\end{figure}

In line with the general trends discussed so far the mass fraction of neutrinos in the total matter leads to a suppression of
the growth of structure as neutrinos free-stream out of the potential wells. The number counts of the most massive objects are therefore
affected as shown in Figure~\ref{f:dhmf}. Within the range of neutrino masses explored in our study we find that the number
count of very massive clusters above 10$^{15} M_\odot/h$ varies by up to a factor of two. These results are in good
qualitative agreement with the ones found previously in \cite{Castorina:2015bma}.

\section{Conclusion}
\label{sec:conclusions}

In this paper we present results from a suite of cosmological N-body simulations with massive neutrinos, carried out with
our relativistic particle-mesh code \textit{gevolution}. This novel approach allows us for the first time to perform such simulations fully
self-consistently within the framework of general relativity, extending in particular to aspects of gauge choice and setting of initial
conditions. The N-body phase space is evolved correctly using a weak-field description that is valid for arbitrary momenta, and hence
we are able to take into account the relativistic nature of neutrino particles. Furthermore, the approach also allows us to extract the additional five gravitational degrees of freedom beyond the Newtonian potential.

We explore six different scenarios for the neutrino mass eigenstates, including as reference the massless case often assumed in
standard cosmology. For the runs with massive neutrinos we consider $\sum m_\nu =$ 0.06~eV (a scenario close to the lowest mass
compatible with neutrino flavour oscillations as seen by particle physics experiments), $\sum m_\nu =$ 0.1~eV (two cases, representing
normal and inverted mass hierarchy, respectively), $\sum m_\nu =$ 0.2~eV, and $\sum m_\nu =$ 0.3~eV (close to the highest mass compatible
with cosmological constraints). At the level of the power spectra of total matter and neutrinos our relativistic results at low redshift
generally show good agreement with those obtained with Newtonian methods, at least when we consider scales deep inside the cosmological
horizon. More specifically, the nonlinear recipes provided by HALOFIT and the latest revision of the Cosmic Emulator, which are both
calibrated to Newtonian simulations, are confirmed to perform to specifications for the cases explored in this paper. While the Cosmic Emulator achieves a better absolute
accuracy on the total matter power spectrum, we find that HALOFIT is still better suited when studying the relative suppression
of power in the presence of massive neutrinos. This is a consequence of the way the error depends on cosmology in the two recipes.

At fixed total matter density, a larger sum of neutrino masses implies a stronger suppression of power at small scales, as most of
the neutrino particles have sufficient thermal energy to stream out of potential wells. Nonlinearities tend to amplify the effect
initially, roughly until most of the structure has virialized. For a minimal mass scenario with $\sum m_\nu =$ 0.06~eV we find
a $\sim$5\% suppression of total matter power compared to the case of massless neutrinos around $k\simeq$ 1~$h$/Mpc where the
suppression is strongest. The size of the effect increases with neutrino mass, reaching almost 25\% for $\sum m_\nu =$ 0.3~eV.
We find that HALOFIT provides a reasonable model for the relative suppression of power, but tends to underestimate the amount of
suppression at intermediate scales 0.1~$h$/Mpc $\lesssim k \lesssim$ 1~$h$/Mpc to the effect of changing the result by up to $\sim$1\%.

The general trend of small scale power suppression also carries forward to the power spectra of the relativistic metric perturbations,
i.e.\ the gravitational slip, the frame-dragging potential, and the tensor perturbation. On large scales the behaviour can be understood
from a second-order calculation. There the sources are quadratic in the matter perturbations and therefore the suppression
effect is seen to be roughly twice as strong as in the matter power spectrum itself. In addition, neutrinos produce some anisotropic
stress already at linear order, dominating the gravitational slip at high redshift and very large scales. 
We show that the stress-energy tensor of our neutrino N-body ensemble fully includes this effect. However, the overall amplitude of the
relativistic perturbations remains very small, so that it will be very difficult to use them for observational constraints.

We finally also study the halo mass function and are able to confirm that it shows a strong dependence on the neutrino mass
at very high halo masses $\gtrsim$~10$^{15}~M_\odot/h$. This can provide a means to get additional constraints on the neutrino mass if it is possible to obtain reliable number counts for such rare, extremely massive objects.
The neutrino mass hierarchy, on the other hand, seems to have little impact on the halo mass function.

In the future it would be interesting to use our numerical approach to study the regime around the free-streaming scale of neutrinos in
more detail. Newtonian simulations are expected to have some subtle issues there because their neutrino propagation violates causality.
Even if this is rectified (e.g.\ according to our recipe given in Section~\ref{sec:newtonian}) there remain the relativistic effects
due to gauge choice and the presence of radiation perturbations that are difficult to deal with in a Newtonian scheme. Future
observations at high redshifts, e.g.\ 21cm surveys~\cite{Kulkarni:2016izh} at $z$ up to 10 will potentially be sensitive to such issues.
The latest version 1.1 of our relativistic code \textit{gevolution}, available at
\url{https://github.com/gevolution-code/gevolution-1.1.git}, provides a self-consistent treatment of all these aspects.

\acknowledgments

We thank K.\ Heitmann, J.\ Lesgourgues, V.\ Poulin, T.\ Tram and M.\ Viel for valuable input, and D.\ Daverio for continued
technical support related to the \lf\ library \cite{David:2015eya} and its interface with \textit{gevolution}.
This work was supported by a grant from the Swiss National Supercomputing Centre (CSCS) under project ID s710. RD and MK also received support from the Swiss National Science Foundation.

\paragraph{} {\small This is an author-created, un-copyedited version of an article published in the Journal of Cosmology and Astroparticle Physics (JCAP). IOP Publishing Ltd is not responsible for any errors or omissions in this version of the manuscript or any version 
derived from it. The Version of Record is available online at \url{https://doi.org/10.1088/1475-7516/2017/11/004}.}

\appendix

\section{Leapfrog integrator for relativistic particles}
\label{app:integrator}

Symplectic integrators \cite{Sanz-Serna:1992} play an important role in the context of N-body simulations.
They are defined by the property to preserve the form $\dd\mathbf{x}\wedge\dd\mathbf{q}$ exactly,
as demanded by Hamiltonian time evolution (in our notation the canonical coordinate is $x^i$ while $q_i$ is the canonical momentum).
In our relativistic weak-field setting, the one-particle Hamiltonian giving rise to the Hamiltonian equations of motion (\ref{eq:geodesic})
and (\ref{eq:velocity}) reads
\begin{multline}
 H(x^i,q_i,\tau) = \sqrt{q^2 + m^2 a^2(\tau)} \left[1 - \chi(x^i,\tau)\right] + \frac{2 q^2 + m^2 a^2(\tau)}{\sqrt{q^2 + m^2 a^2(\tau)}} \phi(x^i,\tau)\\
 + q^j B_j(x^i,\tau) - \frac{1}{2} \frac{q^j q^k h_{jk}(x^i,\tau)}{\sqrt{q^2 + m^2 a^2(\tau)}} \, .
\end{multline}
As explained in Section \ref{sec:numerics}, the last term is expected to be extremely small and we therefore neglect it from now on,
even though it can easily be included. The solution to the equations of motion are computed numerically at discrete time steps.
To illustrate this, let us first consider the limit of non-relativistic motion, $q^2 \ll m^2 a^2$, assuming also the post-Newtonian
ordering of weak fields $|\phi| \ll 1$, $|\chi| \ll |\phi|$, $\sqrt{B^2} \ll |\phi|$. In this approximation the Hamiltonian becomes
\begin{equation}
 H(x^i,q_i,\tau) = m a(\tau) \left[1 + \phi(x^i,\tau)\right] + \frac{q^2}{2 m a(\tau)} \, .
\end{equation}
In this case, the Hamiltonian equations of motion are the Newtonian ones, eq.~(\ref{eq:Ngeodesic}) and (\ref{eq:Nvelocity}).

We now introduce an integer label $\mathbf{n}$ that enumerates the discrete time steps $\tau_\mathbf{n}$, and we use half-integers
to denote midpoints between time steps. A kick-drift-kick scheme for the Newtonian equations reads
\begin{subequations}
\begin{eqnarray}
 q_i^\mathbf{n+\frac{1}{2}} &=& q_i^\mathbf{n} - \frac{d\tau}{2} m a(\tau_\mathbf{n}) \phi_{,i}(x^i_\mathbf{n}, \tau_\mathbf{n})\, ,\label{eq:Nkick1}\\
 x^i_\mathbf{n+1} &=& x^i_\mathbf{n} + \delta^{ij} d\tau \frac{q_j^\mathbf{n+\frac{1}{2}}}{m a(\tau_\mathbf{n+\frac{1}{2}})}\, ,\label{eq:Ndrift}\\
 q_i^\mathbf{n+1} &=& q_i^\mathbf{n+\frac{1}{2}} - \frac{d\tau}{2} m a(\tau_\mathbf{n+1}) \phi_{,i}(x^i_\mathbf{n+1}, \tau_\mathbf{n+1})\, ,\label{eq:Nkick2}
\end{eqnarray}
\end{subequations}
where we introduced the shorthands $q_i^\mathbf{n} \doteq q_i(\tau_\mathbf{n})$ and $x^i_\mathbf{n} \doteq x^i(\tau_\mathbf{n})$. As one
can show by direct computation, this update sequence is symplectic. In practice one can avoid the explicit computation of $q_i$ at
 integer time steps by combining eq.~(\ref{eq:Nkick1}) with eq.~(\ref{eq:Nkick2}) from the previous time step. The resulting
abridged sequence is what we call the leapfrog scheme. We may still call this scheme symplectic even though this property is somewhat
implicit.

Let us now consider the relativistic system of equations with $q^2$ arbitrary. In \textit{gevolution} we use a straightforward
generalization of the above leapfrog scheme, which reads
\begin{subequations}
\begin{multline}
 q_i^\mathbf{n+\frac{1}{2}} = q_i^\mathbf{n} - \frac{d\tau}{2} \Biggl[\frac{2 q^2_\mathbf{n-\frac{1}{2}} + m^2 a^2(\tau_\mathbf{n})}{\sqrt{q^2_\mathbf{n-\frac{1}{2}} + m^2 a^2(\tau_\mathbf{n})}} \phi_{,i}(x^i_\mathbf{n}, \tau_\mathbf{n})\Biggr.\\ \Biggl. - \sqrt{q^2_\mathbf{n-\frac{1}{2}} + m^2 a^2(\tau_\mathbf{n})} \chi_{,i}(x^i_\mathbf{n}, \tau_\mathbf{n}) - \delta^{jk} q_j^\mathbf{n-\frac{1}{2}} B_{k,i}(x^i_\mathbf{n}, \tau_\mathbf{n})\Biggr]\, ,\label{eq:kick1}
\end{multline}
\begin{multline}
 x^i_\mathbf{n+1} = x^i_\mathbf{n} + \delta^{ij} d\tau \frac{q_j^\mathbf{n+\frac{1}{2}}}{\sqrt{q^2_\mathbf{n+\frac{1}{2}} + m^2 a^2(\tau_\mathbf{n+\frac{1}{2}})}} \Biggl[1 + \frac{2 q^2_\mathbf{n+\frac{1}{2}} + 3 m^2 a^2(\tau_\mathbf{n+\frac{1}{2}})}{q^2_\mathbf{n+\frac{1}{2}} + m^2 a^2(\tau_\mathbf{n+\frac{1}{2}})} \phi(x^i_\mathbf{n}, \tau_\mathbf{n}) - \chi(x^i_\mathbf{n}, \tau_\mathbf{n})\Biggr]\\ + \delta^{ij} d\tau B_j(x^i_\mathbf{n}, \tau_\mathbf{n})\, ,\label{eq:drift}
\end{multline}
\begin{multline}
 q_i^\mathbf{n+1} = q_i^\mathbf{n+\frac{1}{2}} - \frac{d\tau}{2} \Biggl[\frac{2 q^2_\mathbf{n+\frac{1}{2}} + m^2 a^2(\tau_\mathbf{n+1})}{\sqrt{q^2_\mathbf{n+\frac{1}{2}} + m^2 a^2(\tau_\mathbf{n+1})}} \phi_{,i}(x^i_\mathbf{n+1}, \tau_\mathbf{n+1})\Biggr.\\ \Biggl.\qquad - \sqrt{q^2_\mathbf{n+\frac{1}{2}} + m^2 a^2(\tau_\mathbf{n+1})} \chi_{,i}(x^i_\mathbf{n+1}, \tau_\mathbf{n+1}) - \delta^{jk} q_j^\mathbf{n+\frac{1}{2}} B_{k,i}(x^i_\mathbf{n+1}, \tau_\mathbf{n+1})\Biggr]\, .\label{eq:kick2}
\end{multline}
\end{subequations}
Note that the first and last step are constructed in a way that we can again avoid an explicit computation of $q_i$ at the integer
time step. This is possible because $q_i$ is time-independent at the background level and can therefore be evaluated at any time
for the purpose of constructing the prefactors of weak-field terms. We remind the reader that we only go to first order in the
gravitational fields for the purpose of solving the equations of motion of the particles.

The scheme defined by eqs.~(\ref{eq:kick1})--(\ref{eq:kick2}) is not exactly symplectic, yet symplecticity is recovered in the limit
of small velocities and post-Newtonian ordering of the weak fields. This is sufficient for the purpose of this work. However, with
a simple adjustment of the drift step it is possible to make the scheme symplectic at leading weak-field order even for relativistic
velocities. All we need to do is replace eq.~(\ref{eq:drift}) by its second-order version,
\begin{multline}
 x^i_\mathbf{n+1} = x^i_\mathbf{n} + \delta^{ij} \frac{d\tau}{2} \frac{q_j^\mathbf{n+\frac{1}{2}}}{\sqrt{q^2_\mathbf{n+\frac{1}{2}} + m^2 a^2(\tau_\mathbf{n})}} \Biggl[1 + \frac{2 q^2_\mathbf{n+\frac{1}{2}} + 3 m^2 a^2(\tau_\mathbf{n})}{q^2_\mathbf{n+\frac{1}{2}} + m^2 a^2(\tau_\mathbf{n})} \phi(x^i_\mathbf{n}, \tau_\mathbf{n}) - \chi(x^i_\mathbf{n}, \tau_\mathbf{n})\Biggr]\\
 + \delta^{ij} \frac{d\tau}{2} \frac{q_j^\mathbf{n+\frac{1}{2}}}{\sqrt{q^2_\mathbf{n+\frac{1}{2}} + m^2 a^2(\tau_\mathbf{n+1})}} \Biggl[1 + \frac{2 q^2_\mathbf{n+\frac{1}{2}} + 3 m^2 a^2(\tau_\mathbf{n+1})}{q^2_\mathbf{n+\frac{1}{2}} + m^2 a^2(\tau_\mathbf{n+1})} \phi(\tilde{x}^i_\mathbf{n+1}, \tau_\mathbf{n+1}) - \chi(\tilde{x}^i_\mathbf{n+1}, \tau_\mathbf{n+1})\Biggr]\\
 + \delta^{ij} \frac{d\tau}{2} B_j(x^i_\mathbf{n}, \tau_\mathbf{n})+ \delta^{ij} \frac{d\tau}{2} B_j(\tilde{x}^i_\mathbf{n+1}, \tau_\mathbf{n+1})\, ,
\end{multline}
where
\begin{equation}
 \tilde{x}^i_\mathbf{n+1} \doteq x^i_\mathbf{n} + \delta^{ij} \frac{d\tau}{2} \frac{q_j^\mathbf{n+\frac{1}{2}}}{\sqrt{q^2_\mathbf{n+\frac{1}{2}} + m^2 a^2(\tau_\mathbf{n})}} + \delta^{ij} \frac{d\tau}{2} \frac{q_j^\mathbf{n+\frac{1}{2}}}{\sqrt{q^2_\mathbf{n+\frac{1}{2}} + m^2 a^2(\tau_\mathbf{n+1})}}
\end{equation}
is the zeroth-order approximation for the new position. A lengthy but straightforward calculation shows that this drift step
fully restores symplecticity at leading weak-field order. Unfortunately the position update now depends on the metric
at final time $\tau_\mathbf{n+1}$ as well, which is usually not available as it will be computed only from the final particle
configuration. This issue can be solved using a predictor-corrector method. If the metric perturbations are time-independent to good
approximation, one may simply use the values from the previous time step. In any case the number of field interpolation operations
will be doubled compared to the simpler drift step (\ref{eq:drift}). For this reason, and since deviations from symplecticity are anyway
small and tend to zero for low velocities, we use the simpler integrator defined by eqs.~(\ref{eq:kick1})--(\ref{eq:kick2}) in our work.

\section{Second-order contributions to the relativistic potentials}
\label{app:2order}

Here we collect the expressions for the convolution integrals that give the second-order contributions to the relativistic
metric perturbations $\chi = \phi - \psi$, $B_i$, and $h_{ij}$. The principles behind this calculation have been laid out in previous
work \cite{Ballesteros:2011cm,Lu:2008ju,Baumann:2007zm} and where necessary we generalize these to the case of multiple matter species.
We also note that the growth rate is no longer independent of scale if massive neutrinos are present, which otherwise allows
for some simplifications. Based on the derivation given in the appendix of \cite{Adamek:2017grt}, the second-order
contribution to the power spectrum of $\chi$ is given by
\begin{multline}
 \Delta^\chi_{(2)}(k) = \frac{1}{2 \pi k^5} \int\! d^3\mathbf{q} \left[T_q^\phi T_{\vert\mathbf{k}-\mathbf{q}\vert}^\phi + 4 \pi G a^2 \sum_i \bar{\rho}_i \frac{T^{\theta,i}_{q}}{q^2}\frac{T^{\theta,i}_{\vert\mathbf{k}-\mathbf{q}\vert}}{\vert\mathbf{k}-\mathbf{q}\vert^2}\right]^2 \!\frac{\Delta^{\zeta,\mathsf{in}}(q)}{q^3} \frac{\Delta^{\zeta,\mathsf{in}}(\vert\mathbf{k}-\mathbf{q}\vert)}{\vert\mathbf{k}-\mathbf{q}\vert^3}\\
 \times \left[3 \left(\mathbf{q}\!\cdot\!\mathbf{k}\right)^2 - 2 k^2 \left(\mathbf{q}\!\cdot\!\mathbf{k}\right) - k^2 q^2\right]^2 \, , 
\end{multline}
where $\Delta^{\zeta,\mathsf{in}}$ is the initial power spectrum of the gauge-invariant curvature perturbation,
and $T^\phi_k$, $T^{\theta,i}_k$ are the linear transfer functions of $\phi$ and the divergence of the velocity for the $i$th
matter species, respectively.

The tensor perturbation $h_{ij}$ follows from the spin-2 projection of the same subset of Einstein's equations, and
hence one obtains a very similar expression,
\begin{multline}
 \Delta^h_{(2)}(k) = \frac{1}{2 \pi k^5} \int\! d^3\mathbf{q} \left[T_q^\phi T_{\vert\mathbf{k}-\mathbf{q}\vert}^\phi + 4 \pi G a^2 \sum_i \bar{\rho}_i \frac{T^{\theta,i}_{q}}{q^2}\frac{T^{\theta,i}_{\vert\mathbf{k}-\mathbf{q}\vert}}{\vert\mathbf{k}-\mathbf{q}\vert^2}\right]^2 \!\frac{\Delta^{\zeta,\mathsf{in}}(q)}{q^3} \frac{\Delta^{\zeta,\mathsf{in}}(\vert\mathbf{k}-\mathbf{q}\vert)}{\vert\mathbf{k}-\mathbf{q}\vert^3}\\
 \times 8\left[\left(\mathbf{q}\!\cdot\!\mathbf{k}\right)^2 - k^2 q^2\right]^2 \, , 
\end{multline}
see appendix B of \cite{Adamek:2014xba} for more details. The only difference is the structure of the term in the second line.

The frame-dragging potential $B_i$ is obtained from the transverse projection of the momentum constraint. At second order the resulting
expression reads
\begin{multline}
 \Delta^B_{(2)}(k) = \frac{4}{\pi k^3} \left(4 \pi G a^2\right)^2 \int\! d^3\mathbf{q} \left[\sum_i \bar{\rho}_i \frac{T^{\theta,i}_q}{q^2} T^{\delta,i}_{\vert\mathbf{k}-\mathbf{q}\vert}\right] \left[\sum_j \bar{\rho}_j \left(\frac{T^{\theta,j}_{\vert\mathbf{k}-\mathbf{q}\vert}}{\vert\mathbf{k}-\mathbf{q}\vert^2} T^{\delta,j}_{q} - \frac{T^{\theta,j}_q}{q^2} T^{\delta,j}_{\vert\mathbf{k}-\mathbf{q}\vert}\right)\right]\\
 \times \frac{\Delta^{\zeta,\mathsf{in}}(q)}{q^3} \frac{\Delta^{\zeta,\mathsf{in}}(\vert\mathbf{k}-\mathbf{q}\vert)}{\vert\mathbf{k}-\mathbf{q}\vert^3} \left[\left(\mathbf{q}\!\cdot\!\mathbf{k}\right)^2 - k^2 q^2\right] \, ,
\end{multline}
where $T^{\delta,i}_k$ are the linear transfer functions for the density perturbations.

\bibliographystyle{JHEP}
\bibliography{gevolution}

\end{document}